\documentclass[a4paper,11pt]{article}
\usepackage{graphicx}
\usepackage[space]{grffile}
\usepackage{latexsym}
\usepackage{textcomp}
\usepackage[normalem]{ulem}
\usepackage{colortbl}
\usepackage{amsfonts,amsmath,amssymb}
\usepackage{natbib}
\usepackage[hidelinks]{hyperref}
\usepackage{subcaption}
\usepackage{etoolbox}
\makeatletter
\usepackage{authblk}
\usepackage[utf8]{inputenc}
\usepackage[english]{babel}
\usepackage[top=0.8in, bottom=0.8in, left=0.8in, right=0.8in]{geometry}
\usepackage{float}
\newcommand*{\doi}[1]{\href{https://doi.org/\detokenize{#1}}{\detokenize{#1}}}

\title{\textbf{Intraseasonal Equatorial Kelvin and Rossby Waves in Modern AI-ML Models}}

\author[1]{Shrutee Jalan}
\author[1,2]{Jai Sukhatme}

\affil[1]{Divecha Centre for Climate Change, Indian Institute of Science, Bengaluru, India}
\affil[2]{Centre for Atmospheric and Oceanic Sciences, Indian Institute of Science, Bengaluru, India}

\date{} 

\begin{document}

\maketitle

\begin{abstract}
\begin{normalsize}
    
\noindent We examine the structure of large-scale convectively coupled Kelvin and Rossby waves in a suite of modern AI-ML models. In particular, multiple runs of PanguWeather, GraphCast, FourCastNet and Aurora are performed to assess the structure of the aforementioned waves. Wavenumber-frequency diagrams of zonal winds from all models show a clear signature of Rossby and Kelvin waves with equivalent depths that are in accord with observations and reanalysis. Composites of Kelvin waves show correct lower and upper troposphere horizontal convergence patterns, vertical tilts in temperature, humidity and vertical velocity as well as the phase relation between temperature and vertical velocity anomalies. Though, differences between models are notable such as smaller vertical tilts and incorrect surface temperature anomalies in GraphCast and relatively weak convergent flows in PanguWeather. The models had much more difficulty with Rossby waves; while the horizontal gyres were captured, the vertical structure of temperature and divergence was incorrect. Apart from unexpected tilts in various fields, the temperature anomaly was inconsistent with the nature of the vertical velocity in all four models. Curiously, moisture and vertical velocity anomalies were much closer to observations. Further, only two models (GraphCast and FourCastNet) captured the simultaneous generation of deep vertical motion with moisture anomalies. In all, while the representation of these large-scale waves is encouraging, issues with the structure of Rossby waves and especially the inconsistency among fields require further investigation. 

\end{normalsize}
\end{abstract}

\makeatletter
\let\@makefnmark\relax
\makeatother
\footnotetext{Both authors contributed equally to this work.}
\footnotetext{Corresponding author: Shrutee Jalan,  \href{mailto:shruteej@iisc.ac.in}{shruteejalan@iisc.ac.in}}


\section{Introduction}

Along with the Madden-Julian Oscillation \citep[MJO;][]{Zhang} and the Boreal Summer Intraseasonal Oscillation \citep[BSISO;][]{kikuchi2021boreal}, convectively coupled equatorial waves \citep[CCEWs;][]{Kiladis-rev} form an integral part of tropical variability at intraseasonal timescales. These waves feature prominently in satellite based observational analyses, such as that of tropical outgoing longwave radiation or brightness temperature \citep{takayabu,wk1999}, as well as in other dynamical fields available from reanalysis \citep{yang2003,hendon-wheeler}. Specifically, Kelvin, Rossby, mixed Rossby-Gravity and Inertia-Gravity waves have been identified and their structure, activity and geographical preference, from both observations and reanalysis, have been carefully documented \citep{WKW,yang2007,naka-taka}. The basic features of these waves are anticipated from dry counterparts in idealized models --- for example, this entire family exists and forms a complete and orthonormal basis in a prototypical shallow water system on an equatorial $\beta$ plane \citep{matsuno1966,Ripa}. In fact, wavenumber-frequency diagrams of turbulent nonlinear numerical simulations of the rotating spherical shallow water system show heightened power along the theoretical dispersion curves of these waves and suggest a natural tendency for a higher fraction of tropical variability to reside in lower frequencies and larger length scales \citep{suhas2020,gar2021,sch2021}.

\noindent In addition to being fundamental modes of the tropical atmosphere, from a practical viewpoint, it has been recognized that organized convection in the tropics, synoptic systems as well as extreme events can develop in an environment shaped by large-scale, low frequency CCEWs \citep{Kiladis-rev}. For example, the modulation of rainfall \citep{lubis,schlueter} and convective systems \citep{cheng2023}, the tendency for cyclone formation associated with Rossby and Kelvin waves \citep{bessafi2006modulation, molinari2007tropical, ventrice2012, schreck} and the occurrence of high intensity rain events \citep{ferrett,latos}. These findings, along with the intraseasonal nature of large-scale CCEWs raise the possibility that a correct representation of these waves in models may give rise to improved subseasonal predictability of synoptic events in the tropical atmosphere \citep{frank-roundy,vitart2018,schreck2020,judt,dias2023}. In fact, it is now recognized that equatorial waves play a role in shaping the overturning regional seasonal Hadley Cell itself \citep{tom-yang,Thakur}.

\noindent Partly motivated by these reasons, the representation of CCEWs in General Circulation Models (GCMs) has been an ongoing subject of inquiry \citep{lin2006,yang2009,straub2010}. This includes the sensitivity of Kelvin waves to convective relaxation time in medium complexity GCMs \citep{Frier}, identification of systematic relationships between intraseasonal variability and model mean states \citep{kimetal2011,wang-li}, connections between a good representation of the MJO and CCEWs \citep{guo}, the nature of Kelvin and Rossby waves in aquaplanet configurations \citep{blackburn,surajit,rios,macdonald}, comparisons of CCEWs in seasonal forecast models with reanalysis \citep{yang2023}, their dependence on moisture entrainment and coupling \citep{peat,suhas2021} and the subseasonal predictability of the waves themselves \citep{janiga}. 

\noindent With the advent of AI-ML based models that have shown an impressive ability for synoptic and medium-range forecasting \citep[for example,][]{pangu2023,GraphCast,FourCastNet,aurora2025}, it is natural to examine if such data-driven models can also be used for longer subseasonal timescale predictability \citep{Weyn,gomez2023,FuXi}. Moreover, given that these models are "learning" the dynamics of the atmosphere from data, a fundamental issue is to examine if they have managed to learn these basic modes of the tropical atmosphere. Experiments in this context were performed by \cite{hakim2024}, where the convergence patterns associated with Kelvin waves and off-equatorial gyres of Rossby waves were noted in response to an imposed tropical heat source in the data driven PanguWeather model \citep{pangu2023}. 
In this paper, we continue this line of inquiry by examining tropical wavenumber-frequency diagrams of four modern AI-ML models and the composite structure of the equatorial waves that develop in these model runs. The structure of the paper is as follows: In Section 2, we briefly describe the models, provide details of the runs performed and the data analysis tools used in the study. Wavenumber-frequency diagrams of these models are presented in Section 3. Section 4 focuses on Kelvin waves and Section 5 on equatorial Rossby waves. 
Finally, Section 6 concludes with a summary and discussion of the results.

\section{Models and Methodology}

This study uses four state-of-the-art data-driven weather models: Pangu-Weather \citep{pangu2023}, GraphCast \citep{GraphCast}, FourCastNet V2 \citep{FourCastNet} and Aurora \citep{aurora2025}. While each of these models has a different machine learning architecture, Pangu, GraphCast and FourCastNet V2 are trained on 39 years of ERA5 reanalysis \citep{hersbach2020era5}, whereas Aurora, being a so-called foundation model, uses a diverse set of data that includes other reanalysis products \citep[such as MERRA,][]{rienecker2011merra}, CMIP climate simulations \citep{eyring2016overview} \textcolor{black}{and is fine-tuned using ECMWF’s high-resolution IFS (HRES) forecasts through 2022}. 
PanguWeather utilizes a 3D Earth-Specific Transformer (3DEST) based on a Swin Transformer encoder–decoder, operating at a resolution of 0.25$^\circ \times$0.25$^\circ$ with 13 vertical levels, including five pressure-level and three surface variables. FourCastNet V2 uses a Spherical Fourier Neural Operator (SFNO) to forecast five variables on 13 pressure levels and five single-level fields, also at 0.25$^\circ \times$0.25$^\circ$ resolution and 6-hour intervals. Aurora integrates a 3D Swin Transformer with Perceiver-style encoders and decoders, producing 6-hourly forecasts from two consecutive timesteps at a finer resolution of 0.1$^\circ \times$0.1$^\circ$, with five pressure level and two single-level variables across 13 vertical levels. GraphCast adopts a graph neural network architecture, implemented here at a coarser resolution configuration (1$^\circ \times$1$^\circ$), also uses two input timesteps and produces 6-hourly forecasts with six pressure-level and five single level variables. 

\noindent In our numerical experiments, each model is initialized on ten randomly selected dates through the year to capture a range of tropical variability and is run freely for four months (the starting dates used are; 2019-05-12, 2019-09-15, 2020-02-05, 2020-07-10, 2021-04-07, 2021-11-11, 2022-03-28, 2022-08-17, 2023-06-03, and 2023-12-22). Rather than focus on particular events, the aim here is to examine the intraseasonal variability shown by free runs of the models. To analyze tropical CCEWs, model outputs are spectrally filtered to isolate specific spatial features associated with different equatorial waves \citep{wk1999}. For models that do not provide specific humidity and pressure velocity as output variables, we derive specific humidity from relative humidity and temperature \citep{bolton1980computation, wallace2006atmospheric}, and compute pressure velocity using the kinematic method \citep{holton}. Specifically, at each pressure level, the pressure velocity is obtained by integrating the continuity equation from the surface to that level, assuming the pressure velocity at the surface is zero.


\section{Wavenumber Frequency Diagrams}

\begin{figure}[H]
\begin{centering}
    \includegraphics[width=0.7\linewidth]{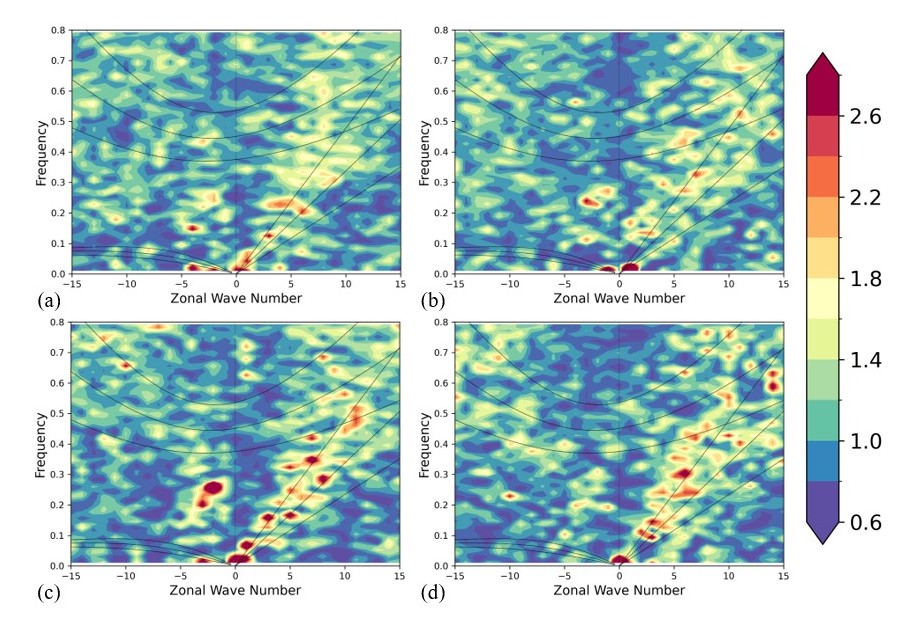}
    \caption{Wavenumber-frequency spectra of symmetric component of the 250 mbar zonal winds for (a) PanguWeather, (b) GraphCast, (c) FourCastNet and (d) Aurora.}
    \label{fig1}
\end{centering}
\end{figure}

As a measure of intraseasonal activity, we first present wavenumber-frequency diagrams constructed from one of the runs of the four AI-ML models. At the outset, the background spectrum \citep[Figure S1 \& S2, as estimated via the NCL routine by progressive smoothening,][]{ncl} is red in character and the 250 mbar wind spectrum has a slight eastward bias, while the 850 mbar wind spectrum has slight westward bias \citep{hendon-wheeler}, though they are largely symmetric in nature. Moreover, the decorrelation timescale of the lower-level winds (Figure S3) is comparable to reanalysis \citep{hendon-wheeler}, except for GraphCast where the winds show much longer persistence especially at large scales. Figure \ref{fig1} shows the spectrum corresponding to the symmetric component of the 250 mbar zonal wind, and comparing with spectra of reanalysis zonal winds \citep{teruya2021}, quite clearly, all four models show significant activity in the bands that correspond to Kelvin waves \citep{wk1999}  --- the three theoretical curves shown here are for equivalent depths of 12 m, 25 m and 50 m. Rossby waves are also identifiable in the spectra, though the wavenumbers with largest power at slightly larger (smaller scales) in Aurora (Figure \ref{fig1}d). In fact, there is a tendency for the accumulation of energy in large scales and low frequencies as is observed in from real-world data \citep{takayabu,wk1999}. While the Rossby waves are predominantly of a quasi-biweekly timescale with length scales less than wavenumber 5, Kelvin waves extend from about the synoptic time scale to 20 days, and up to approximately wavenumber 10. 

\begin{figure}[H]
\begin{centering}
        \includegraphics[width=0.7\linewidth]{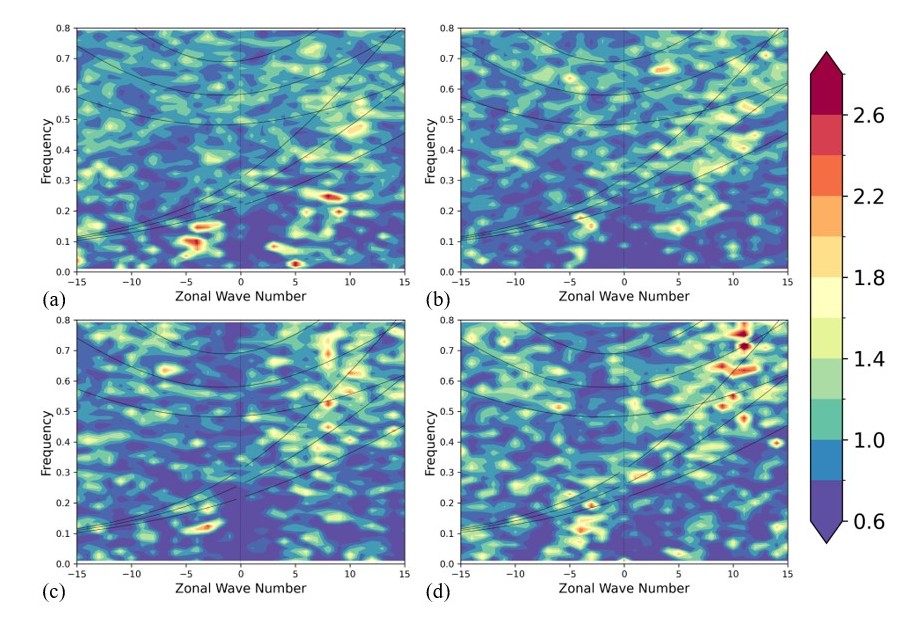}
    \caption{Wavenumber-frequency spectra of antisymmetric component of the 250 mbar zonal winds for (a) PanguWeather, (b) GraphCast, (c) FourCastNet and (d) Aurora.}
    \label{fig2}
\end{centering}
\end{figure}
\noindent We also note a fair amount of power at long time periods in eastward propagating disturbances (outside the Kelvin band), and this may be an indication of the MJO \citep{Zhang}, but our simulations might be too short or do not have enough events to adequately resolve these low frequencies. Moreover, the MJO is seen more clearly in measures of moist convection such as Outgoing Longwave Radiation rather than zonal winds \citep{hendon-wheeler,teruya2021}. On shorter timescales, there is power scattered in the diagrams, but there doesn't appear to be a clear preference for the dispersion curves of the inertia-gravity waves. Some spectral power is also observed for westward-propagating disturbances with periods ranging from 3 to 10 days and zonal wavenumbers between 0 and 5. This is particularly clear in the FourCastNet runs (Figure \ref{fig1}c) and coincides with the signature of global external Rossby waves \citep{hendon-wheeler}.
In a similar manner, spectra of the antisymmetric component of the upper-level zonal wind (Figure \ref{fig2}) also does not show any clear alignment with inertia-gravity and mixed Rossby-gravity waves. 
Signals appear similar in character but notably stronger in the 850 mbar zonal flow (not shown).

\noindent For the rest of the manuscript, we focus on intraseasonal Kelvin and Rossby waves that are clearly identifiable in the wavenumber-frequency plots. 
Based on these diagrams, Kelvin waves are identified by retaining zonal wavenumbers 1–10 and time periods between 10 and 20 days, while Rossby waves are identified using zonal wavenumbers 2–5 and time periods between 12 and 48 days. For Kelvin waves, composite fields are constructed from events which are selected based on the zonal wind variance within the 5N–5S and 135E–160E region. We identify all events in which the zonal wind variance exceeds one standard deviation above the mean climatological variance in this region. These selected events are then used to construct composites of various fields to examine typical structures associated with Kelvin waves. Vertical profiles are computed at the equator. For Rossby waves, activity is measured using geopotential height variance at 250 mb within the 15N-5N and 100E-140E region. Events are identified when geopotential variance exceeds one standard deviation above the climatological mean in this domain. Composite fields are then constructed based on these events to examine the structure associated with Rossby waves and vertical fields are computed at 10N.


\section{Kelvin Waves}

\begin{figure}[H]
    \begin{subfigure}{0.5\textwidth}
        \includegraphics[width=\linewidth]{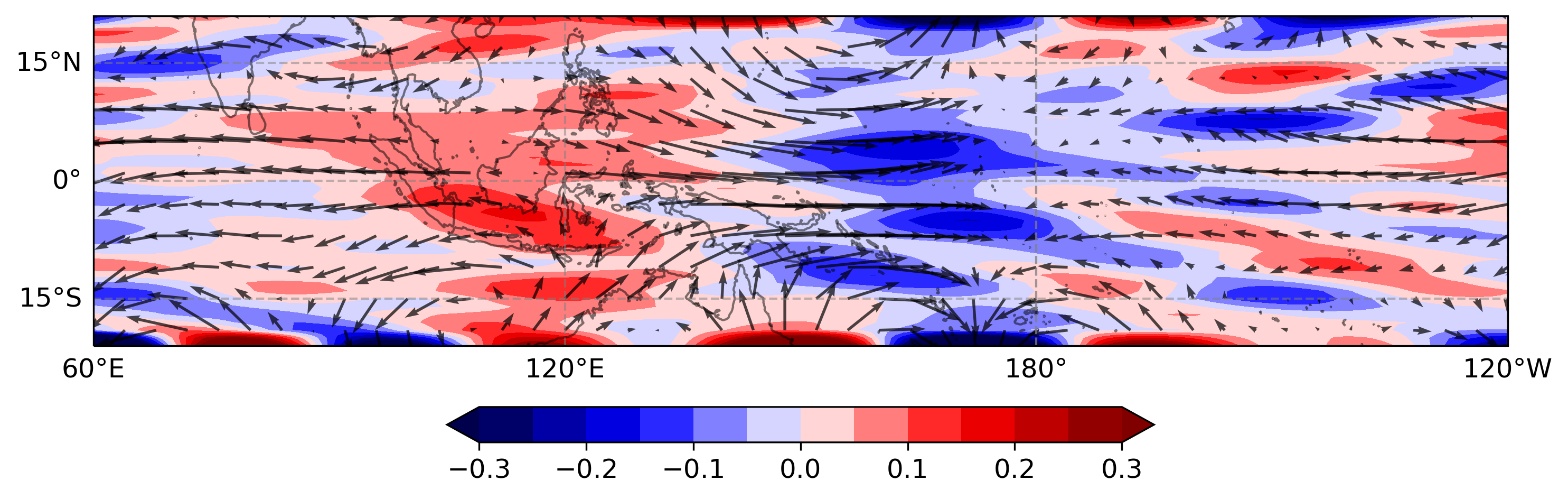}
    \end{subfigure}
    \begin{subfigure}{0.5\textwidth}
        \includegraphics[width=\linewidth]{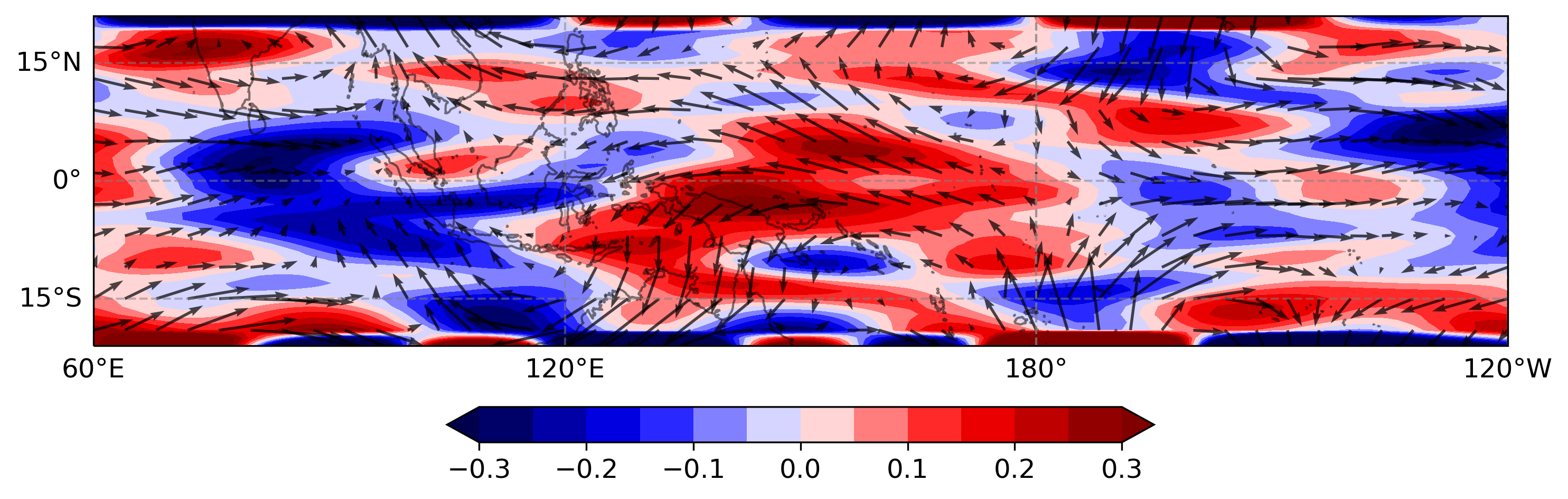}
    \end{subfigure}
    \begin{subfigure}{0.49\textwidth}
        \includegraphics[width=\linewidth]{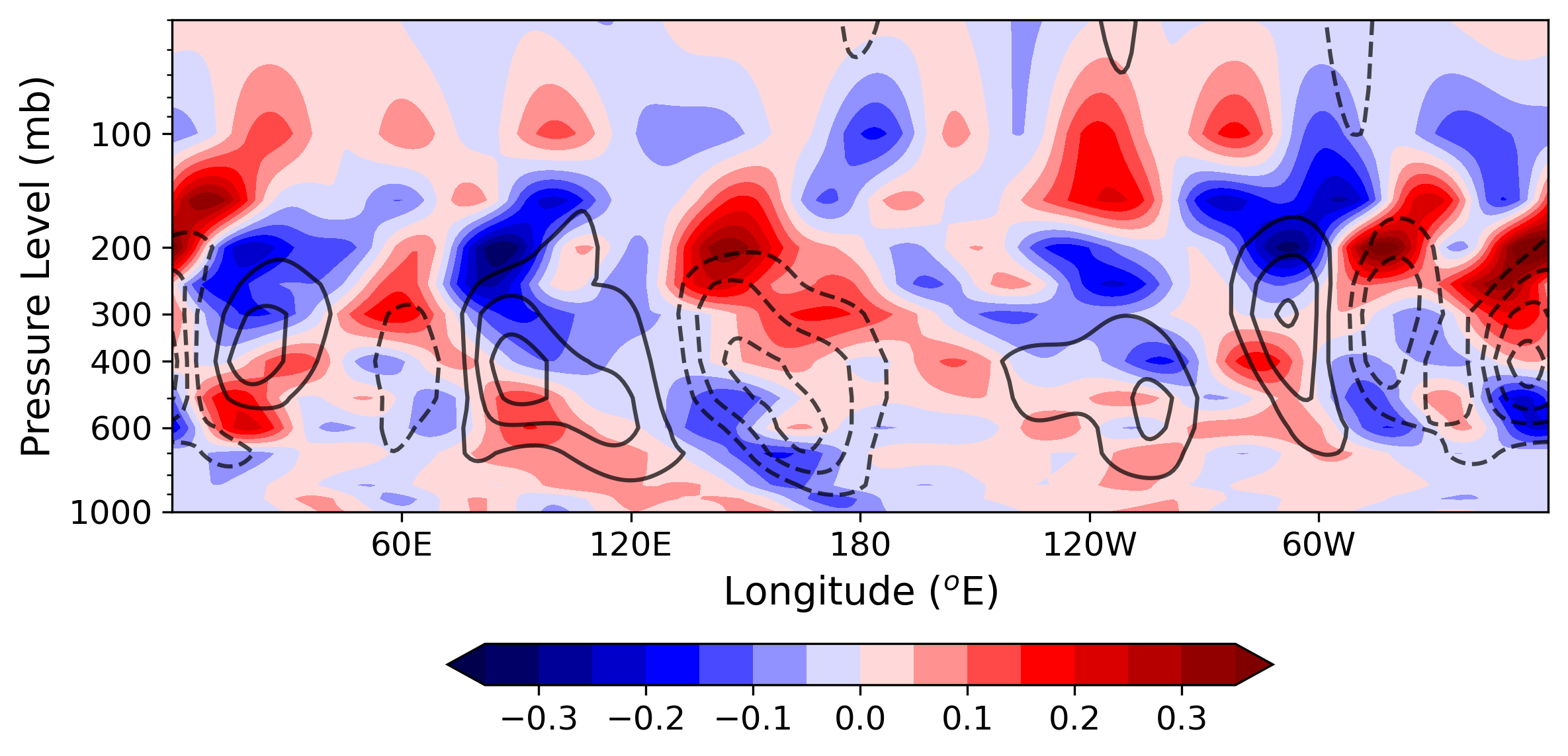}
        \caption{}
    \end{subfigure}
    \begin{subfigure}{0.49\textwidth}
        \includegraphics[width=\linewidth]{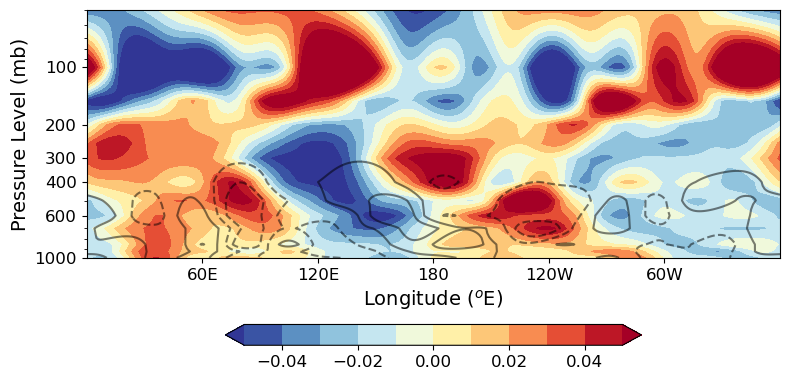}
        \caption{}
    \end{subfigure}
    \caption{Kelvin Wave composite in PanguWeather. Horizontal winds (quivers) and divergence (colours, in $10^{-6}$ $s^{-1}$ ) at (a) 850 mbar, (b) 250 mbar. Vertical profiles of (c) divergence (colours, in $10^{-6}$ $s^{-1}$) with pressure velocity anomaly (contours, in Pa/s solid is positive, dashed is negative), (d) temperature anomalies (colours, in K) with specific humidity anomaly (contours, in gm/kg, solid is positive, dashed is negative).}
    \label{fig3}
\end{figure}

\begin{figure}[H]
    \begin{subfigure}{0.5\textwidth}
        \includegraphics[width=\linewidth]{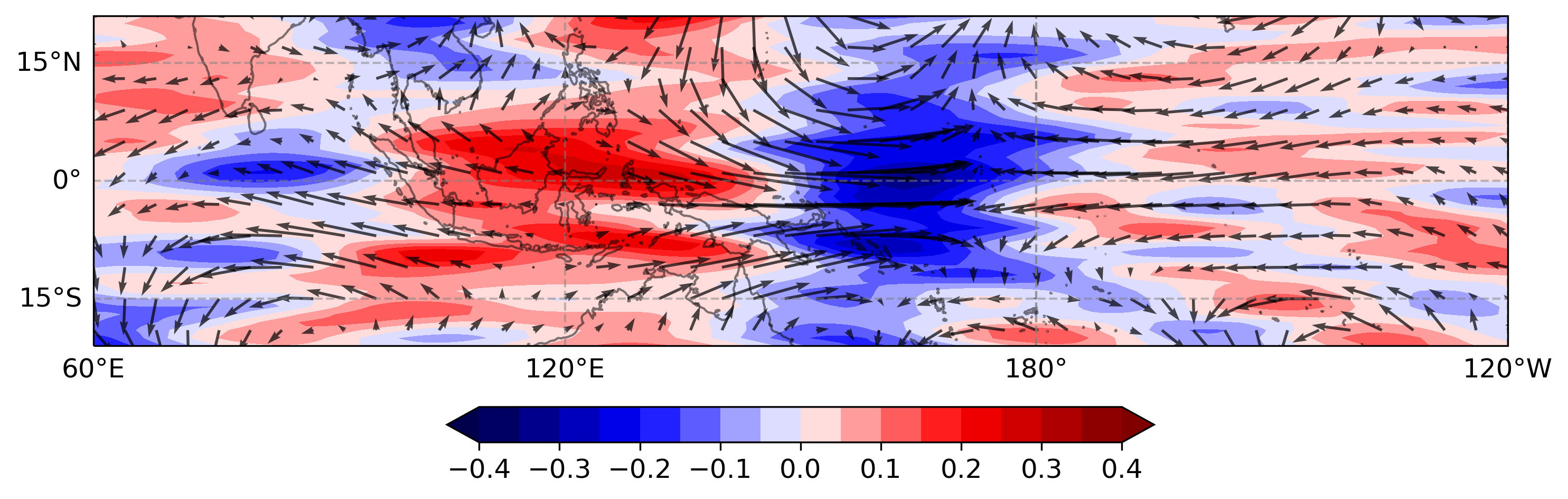}
        \caption{}
    \end{subfigure}
    \begin{subfigure}{0.5\textwidth}
        \includegraphics[width=\linewidth]{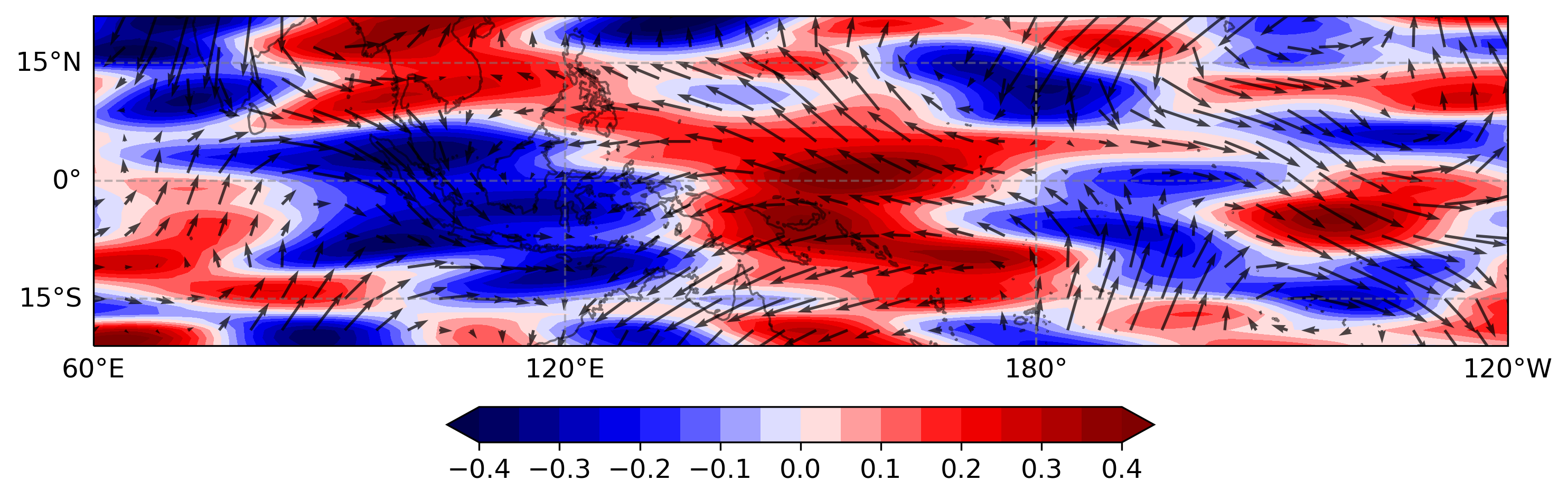}
        \caption{}
    \end{subfigure}
    \begin{subfigure}{0.49\textwidth}
        \includegraphics[width=\linewidth]{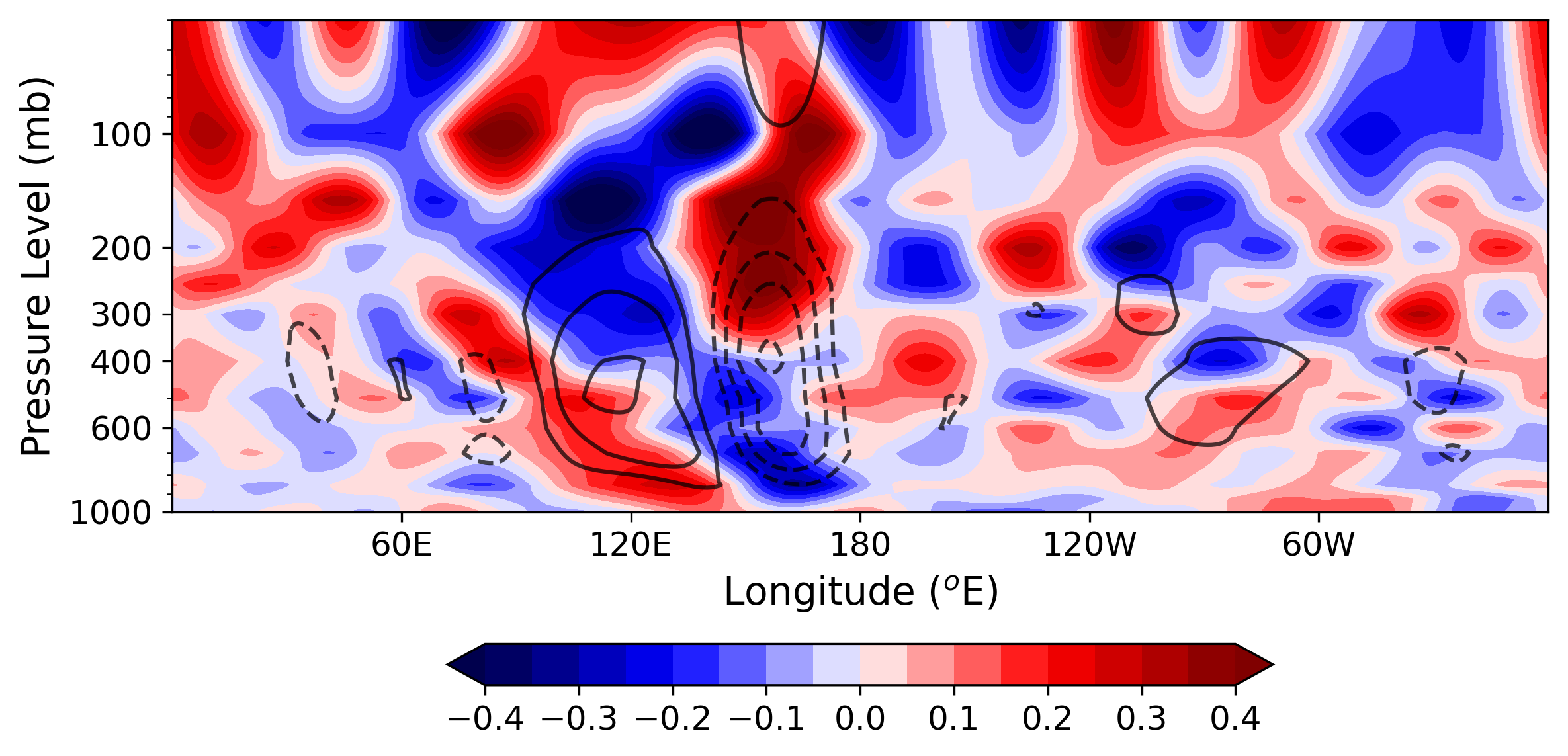}
        \caption{}
    \end{subfigure}
    \begin{subfigure}{0.49\textwidth}
        \includegraphics[width=\linewidth]{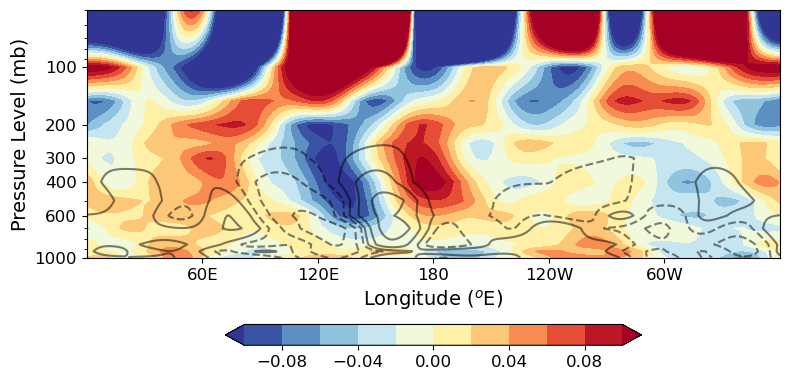}
        \caption{}
    \end{subfigure}
    \caption{Kelvin Wave composite in GraphCast. Horizontal winds (quivers) and divergence (colours, in $10^{-6}$ $s^{-1}$ ) at (a) 850 mbar, (b) 250 mbar. Vertical profiles of (c) divergence (colours, in $10^{-6}$ $s^{-1}$) with pressure velocity anomaly (contours, in Pa/s solid is positive, dashed is negative), (d) temperature anomalies (colours, in K) with specific humidity anomaly (contours, in gm/kg, solid is positive, dashed is negative).}
    \label{fig4}
\end{figure}

\begin{figure}[H]
    \begin{subfigure}{0.5\textwidth}
        \includegraphics[width=\linewidth]{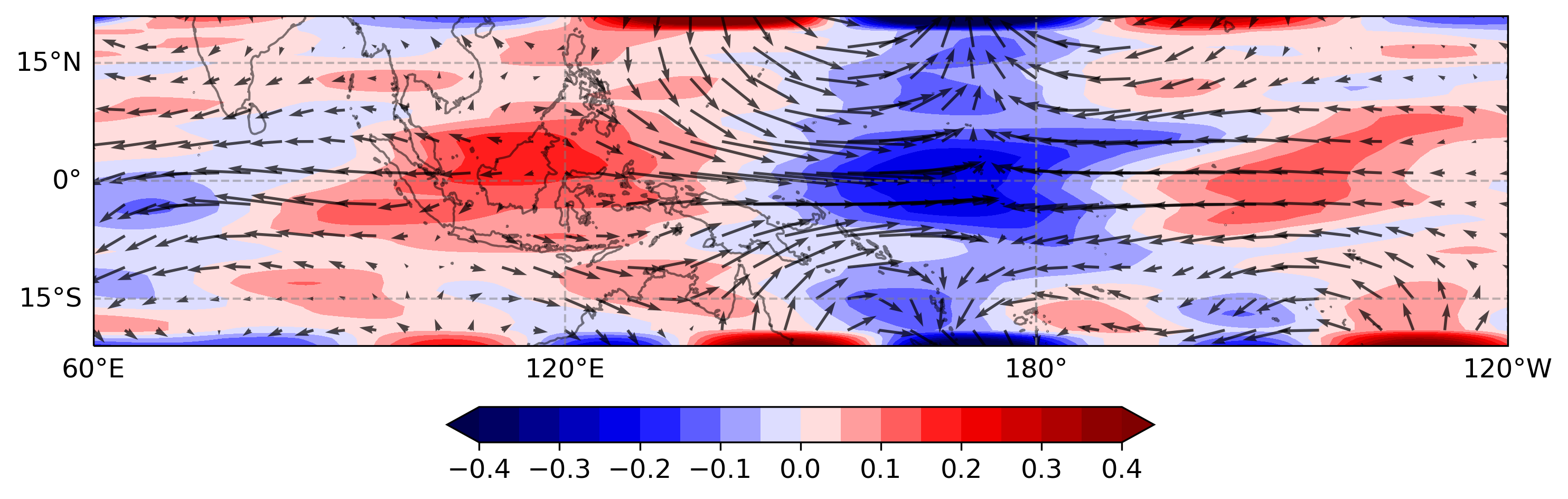}
        \caption{}
    \end{subfigure}
    \begin{subfigure}{0.5\textwidth}
        \includegraphics[width=\linewidth]{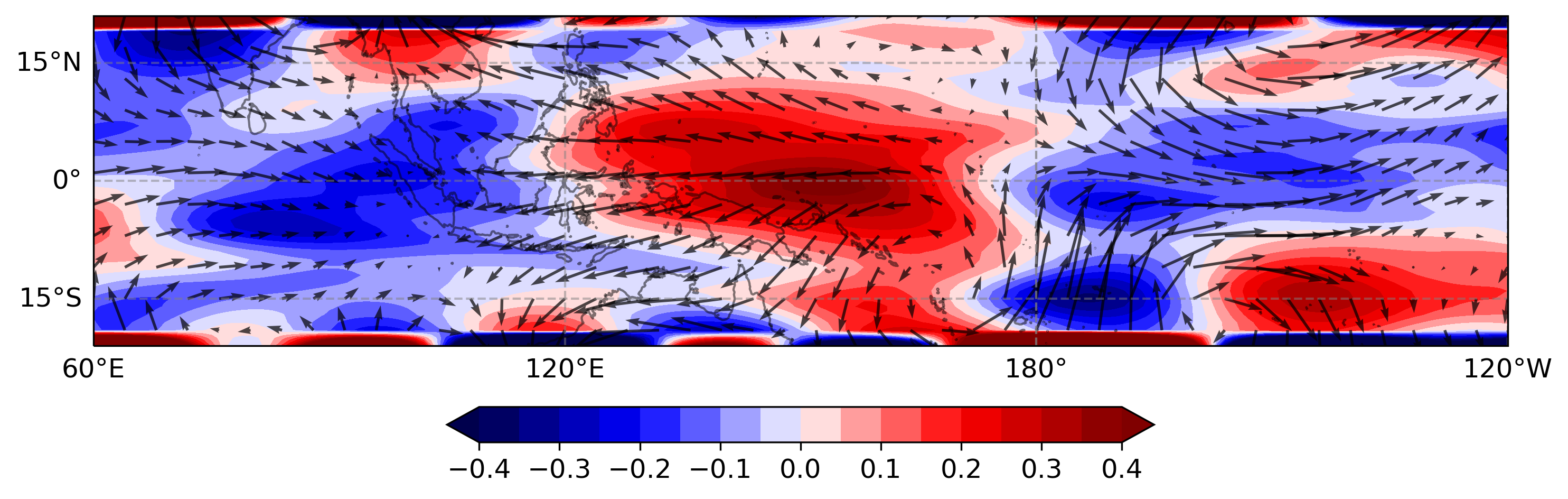}
        \caption{}
    \end{subfigure}
    \begin{subfigure}{0.49\textwidth}
        \includegraphics[width=\linewidth]{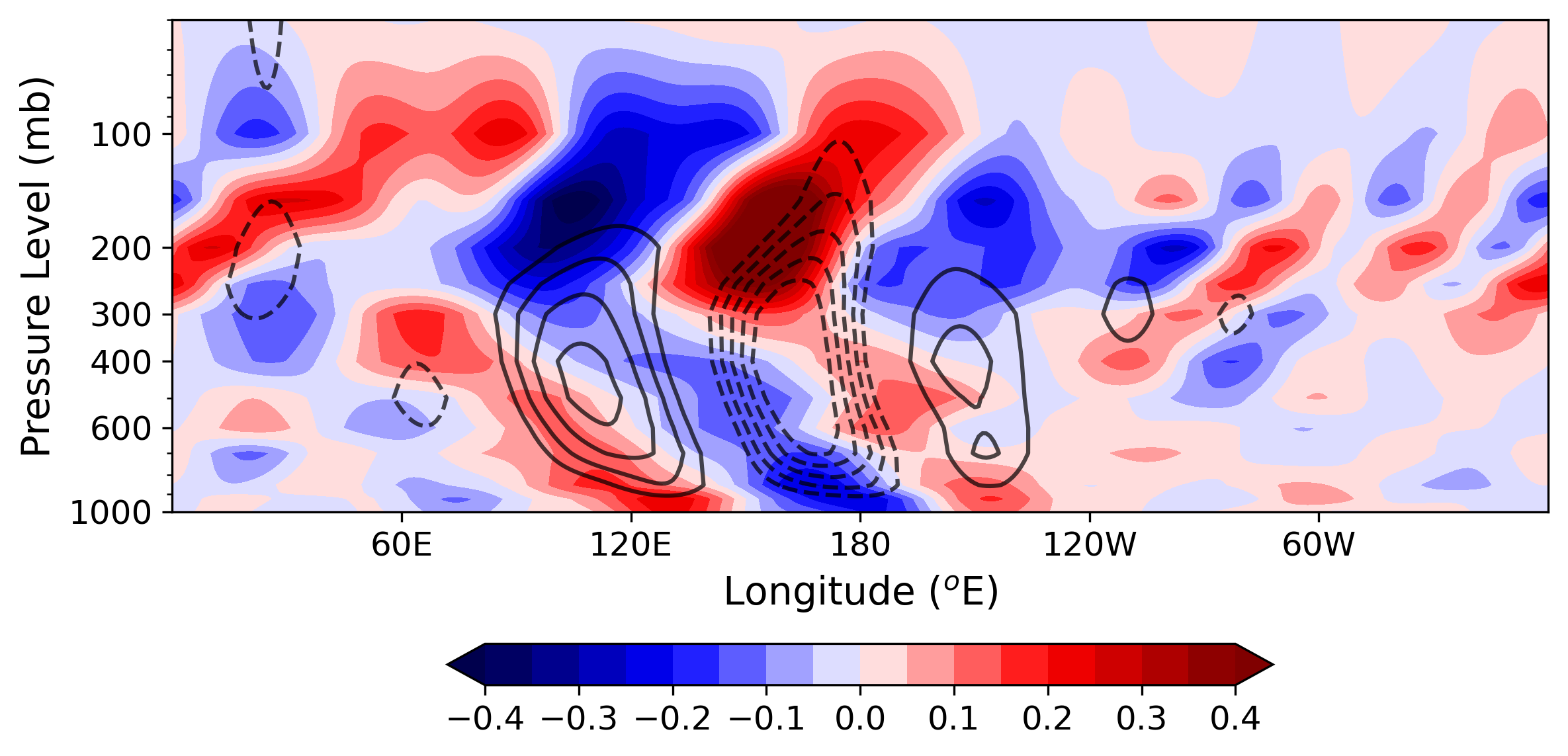}
        \caption{}
        \label{fig:subfig2}
    \end{subfigure}
    \begin{subfigure}{0.49\textwidth}
        \includegraphics[width=\linewidth]{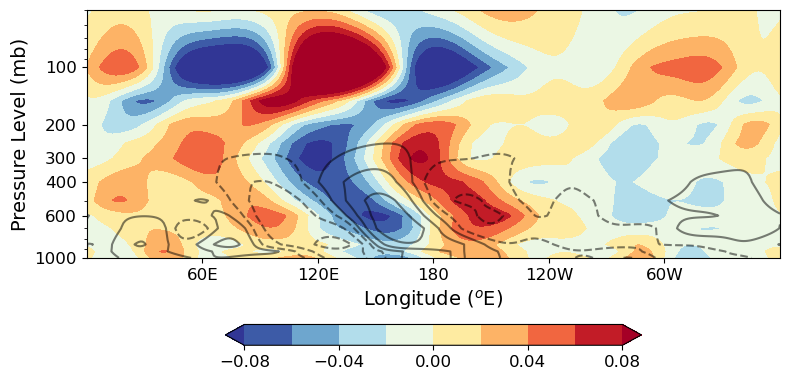}
        \caption{}
    \end{subfigure}
    \caption{Kelvin Wave composite in FourCastNet. Horizontal winds (quivers) and divergence (colours, in $10^{-6}$ $s^{-1}$ ) at (a) 850 mbar, (b) 250 mbar. Vertical profiles of (c) divergence (colours, in $10^{-6}$ $s^{-1}$) with pressure velocity anomaly (contours, in Pa/s solid is positive, dashed is negative), (d) temperature anomalies (colours, in K) with specific humidity anomaly (contours, in gm/kg, solid is positive, dashed is negative).}
    \label{fig5}
\end{figure}

\begin{figure}[H]
    \begin{subfigure}{0.5\textwidth}
        \includegraphics[width=\linewidth]{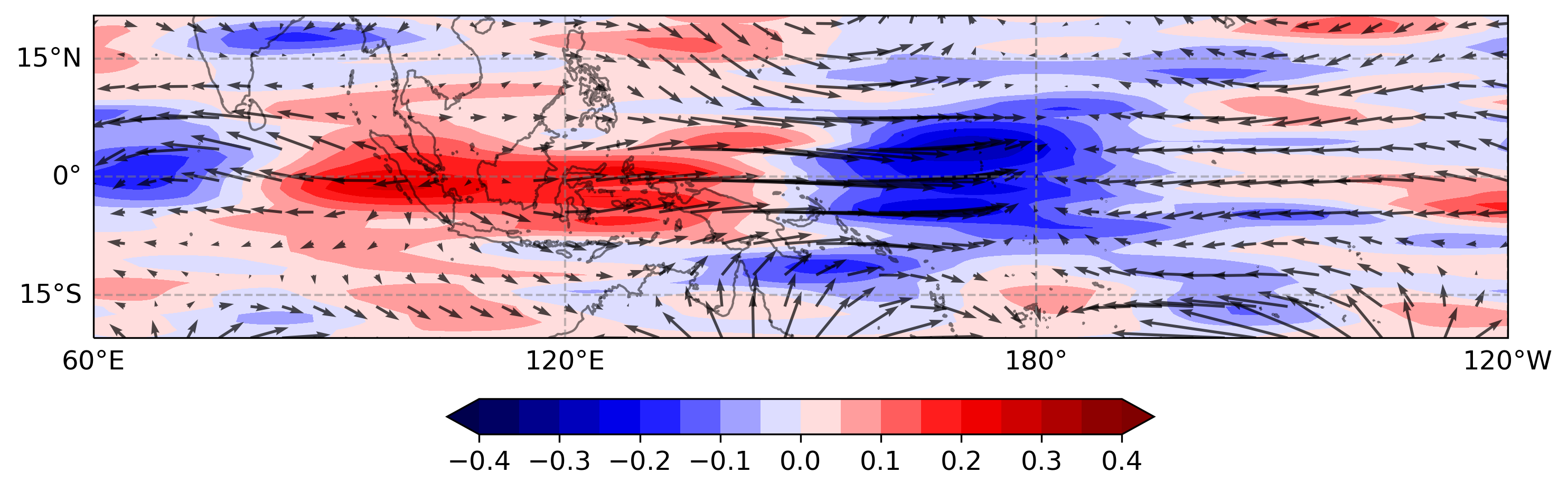}
        \caption{}
    \end{subfigure}
    \begin{subfigure}{0.5\textwidth}
        \includegraphics[width=\linewidth]{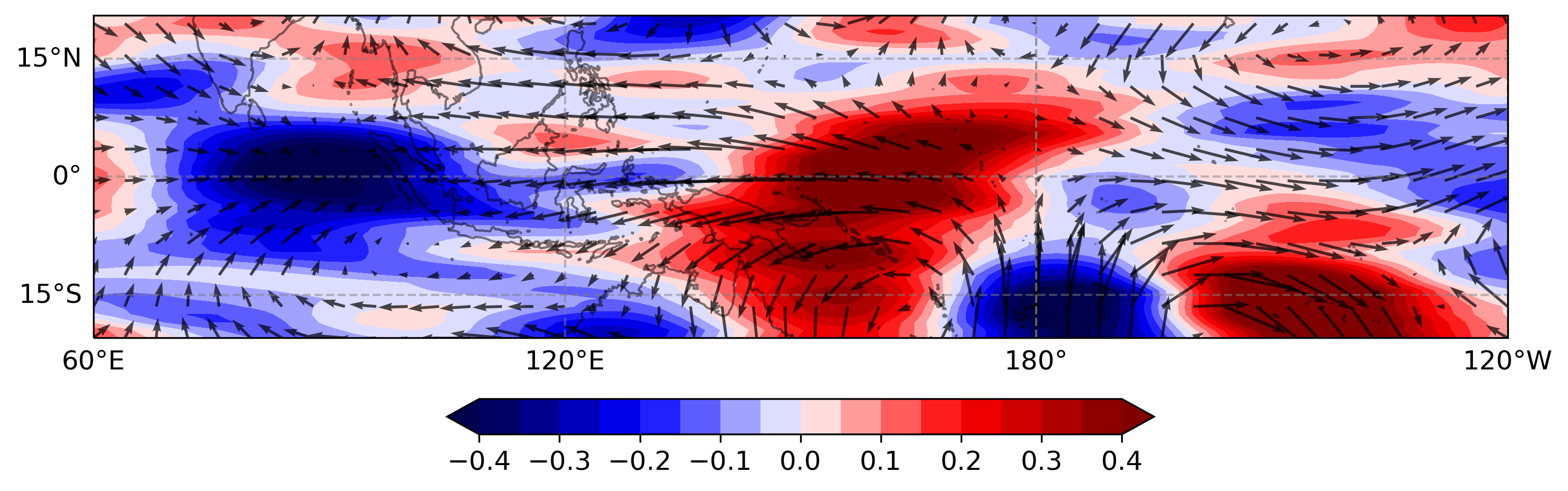}
        \caption{}
    \end{subfigure}
    \begin{subfigure}{0.49\textwidth}
        \includegraphics[width=\linewidth]{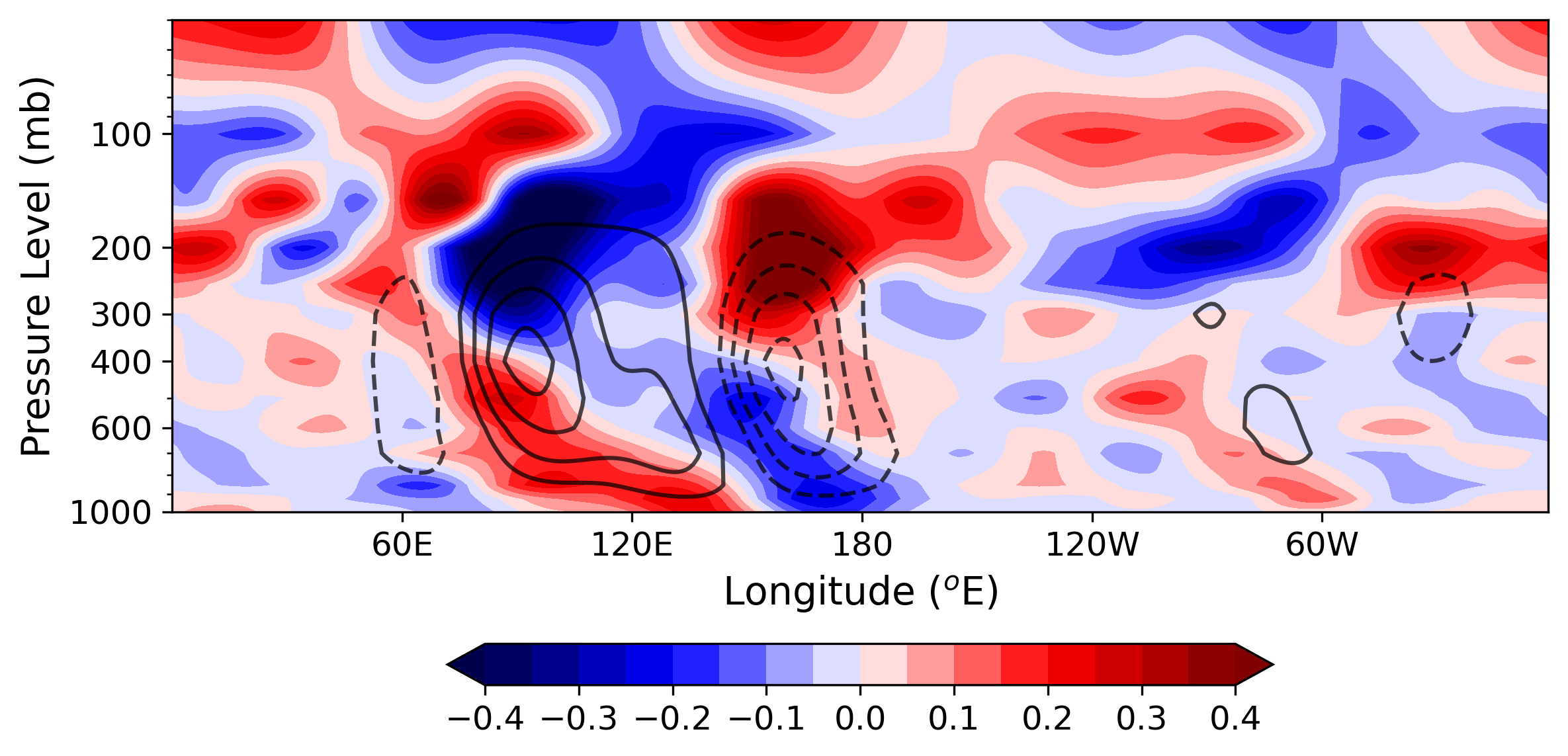}
        \caption{}
    \end{subfigure}
    \begin{subfigure}{0.49\textwidth}
        \includegraphics[width=\linewidth]{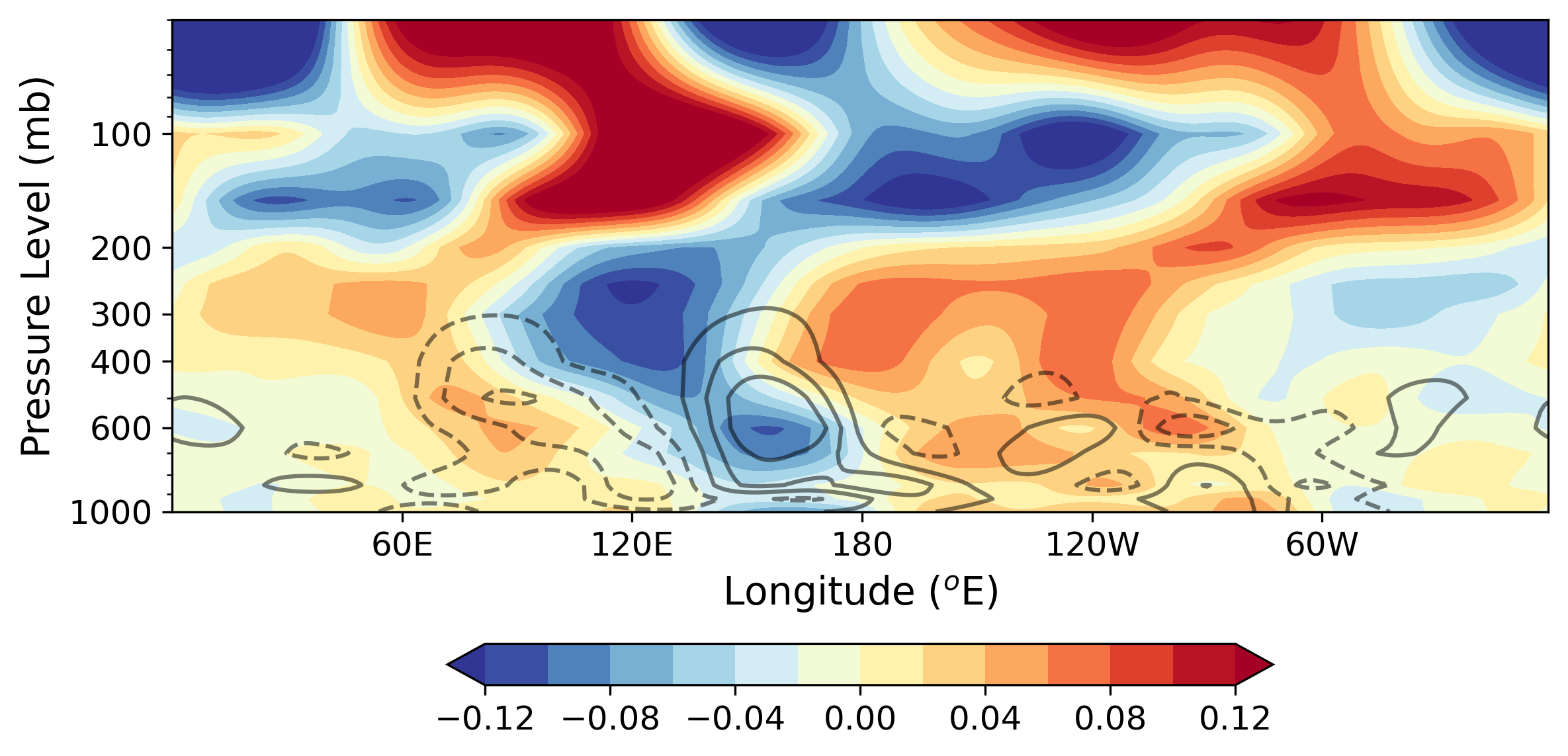}
        \caption{}
    \end{subfigure}
    \caption{Kelvin Wave composite in Aurora. Horizontal winds (quivers) and divergence (colours, in $10^{-6}$ $s^{-1}$ ) at (a) 850 mbar, (b) 250 mbar. Vertical profiles of (c) divergence (colours, in $10^{-6}$ $s^{-1}$) with pressure velocity anomaly (contours, in Pa/s solid is positive, dashed is negative), (d) temperature anomalies (colours, in K) with specific humidity anomaly (contours, in gm/kg, solid is positive, dashed is negative).}
    \label{fig6}
\end{figure}

We examine the structure of the composite of Kelvin waves in each model and compare them with characteristics reported in observations and reanalysis. Maps of the divergence (colors) and horizontal velocity field (quivers) at 850 mbar and 250 mbar in PanguWeather, GraphCast, FourCastNet and Aurora models are shown in Figures \ref{fig3}a,b, \ref{fig4}a,b, \ref{fig5}a,b and \ref{fig6}a,b, respectively. All models exhibit a strong convergence–divergence pattern at 850 mbar. This pattern is associated with pronounced zonal wind anomalies that are nearly symmetric about the equator. At 250 mbar, the pattern reverses, with upper-level convergence (divergence) located to the west of regions of lower-level divergence (convergence), accompanied by a meridional flow transporting mass away from the equator. This basic lower and upper level structure is in accord with the theory \citep{matsuno1966} and observations \citep{straub2003}.

\noindent Vertical profiles of divergence (colors) \& pressure velocity anomalies (contours) and temperature (colors) \& specific humidity anomalies (contours) are shown in panels c,d of Figures \ref{fig3}-\ref{fig6}. Much like reanalysis based composites \citep{naka-taka}, the vertical convergence–divergence structure shows a westward tilt with height up to about 200 mbar, above which an eastward tilt becomes apparent. Consistent with low-level divergence (convergence), positive (negative) pressure velocity anomalies are observed that are largest in the middle troposphere around the 400–600 mbar in all models. The pressure velocity in all models (Figures \ref{fig3}c, \ref{fig4}c, \ref{fig5}c \& \ref{fig6}c) exhibits a slight westward tilt with height up to 200 mbar, resembling the vertical structure of the convergence–divergence field. Temperature anomalies (Figures \ref{fig3}d, \ref{fig4}d, \ref{fig5}d \& \ref{fig6}d) are generally warmer (cooler) just west of regions of low-level divergence (convergence), with peak amplitudes between 700 and 200 mbar. These anomalies also show a westward tilt with height up to 200–300 mbar and an eastward tilt above. All models also capture a strong temperature anomaly signal at and above 100 mbar. Specific humidity anomalies are negative (positive) slightly east of the temperature anomalies and are physically consistent with regions of surface divergence (convergence) and upward (downward) motion. These anomalies are large in the lower to mid-troposphere and exhibit a slight westward tilt with height. 

\noindent The westward tilt of the divergence, vertical velocity, specific humidity and temperature anomalies is consistent with observations and reanalysis \citep{straub2003,naka-taka}, though the level of maximum vertical velocity is displaced somewhat downward in the AI-ML model composites. The cold over warm structure to the east, the coherent upper tropospheric temperature anomalies and the relatively weaker tilt of the vertical velocity as compared to the temperature and specific humidity fields are also in accord with Kelvin waves in reanalysis and observations \citep{straub2003,Kiladis-rev,naka-taka}.

\noindent Although the overall composite structure is identifiable as a Kelvin wave in all the models, there are differences in how clearly and accurately they are represented. While the 850 mbar and 250 mbar field patterns are well captured by all models, wind strength is weakest in PanguWeather (Figure \ref{fig3}a,b). In terms of vertical velocity profile, PanguWeather (Figure \ref{fig3}c) appears the least organized, whereas GraphCast (Figure \ref{fig4}c) lacks the clearly defined vertical tilt and is more upright in character than the rest of the models. In contrast, both Aurora (Figure \ref{fig6}c) and FourCastNet (Figure \ref{fig5}c) capture the expected tilt more clearly. Aurora shows a westward tilt up to about 400 mbar, above which the composite slightly tilts east. FourCastNet represents the vertical velocity structure most clearly, with a pronounced westward tilt up to nearly 200 mbar and an eastward tilt at higher levels. For vertical temperature profiles, PanguWeather (Figure \ref{fig3}d) is characterized by a weaker magnitudes of anomalies, particularly in the lower troposphere. Although it captures the overall structure, the pattern is not as well defined as the other models. In GraphCast (Figure \ref{fig4}d), the near-surface cold anomalies are inaccurately represented, with warmer temperatures appearing in their place. Above 700 mb, however, the temperature anomaly structure aligns more closely with expectations. In contrast, both FourCastNet (Figure \ref{fig5}d) and Aurora (Figure \ref{fig6}d) produce well-defined and coherent temperature anomaly structures, including more accurate representation of near-surface anomalies. A peculiar feature observed in GraphCast and Aurora is the presence of strong alternating temperature anomalies at top of atmosphere. With respect to specific humidity profiles, PanguWeather, GraphCast, and Aurora struggle to capture well-defined anomalies in the lower troposphere, particularly below 800 mb. In this regard, FourCastNet (Figure \ref{fig5}d) seems to be the most coherent of the models. 
\vspace{0.5cm}

\begin{figure}[H]
    \begin{subfigure}{0.25\textwidth}
        \includegraphics[width=\linewidth]{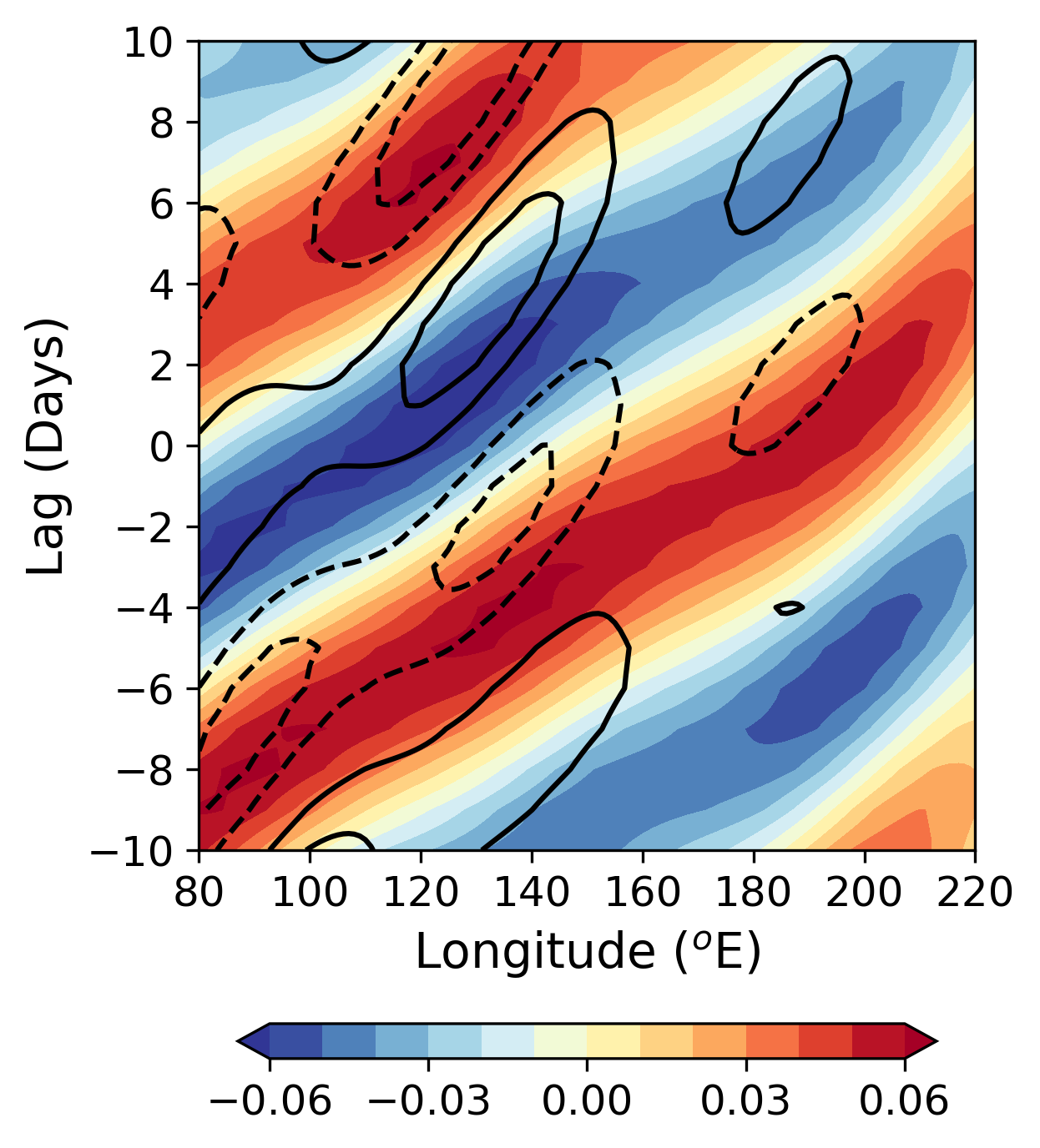}
        \caption{}
    \end{subfigure}
    \hspace{-0.5em}
    \begin{subfigure}{0.25\textwidth}
        \includegraphics[width=\linewidth]{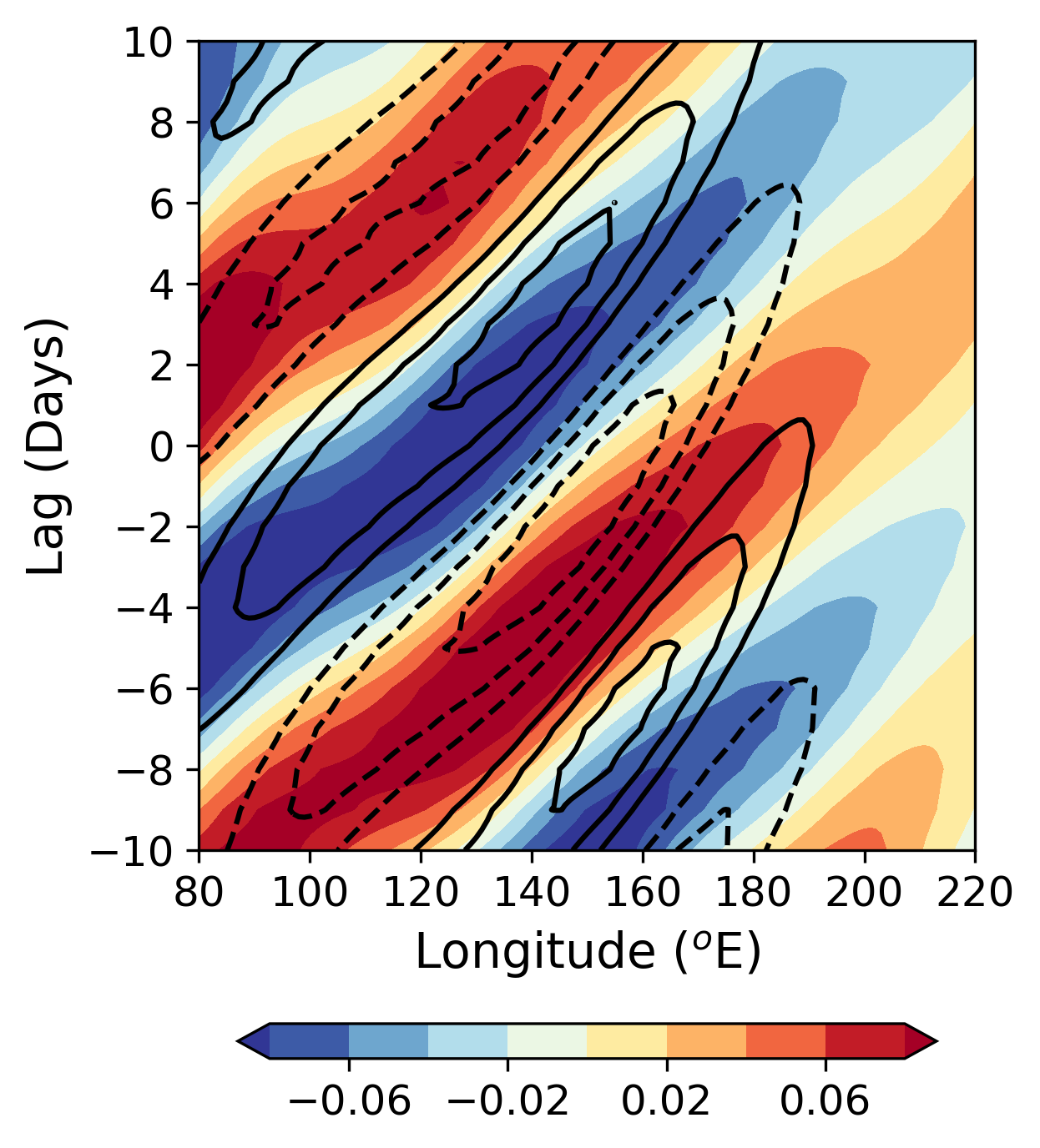}
        \caption{}
    \end{subfigure}
    \hspace{-0.5em}
    \begin{subfigure}{0.25\textwidth}
        \includegraphics[width=\linewidth]{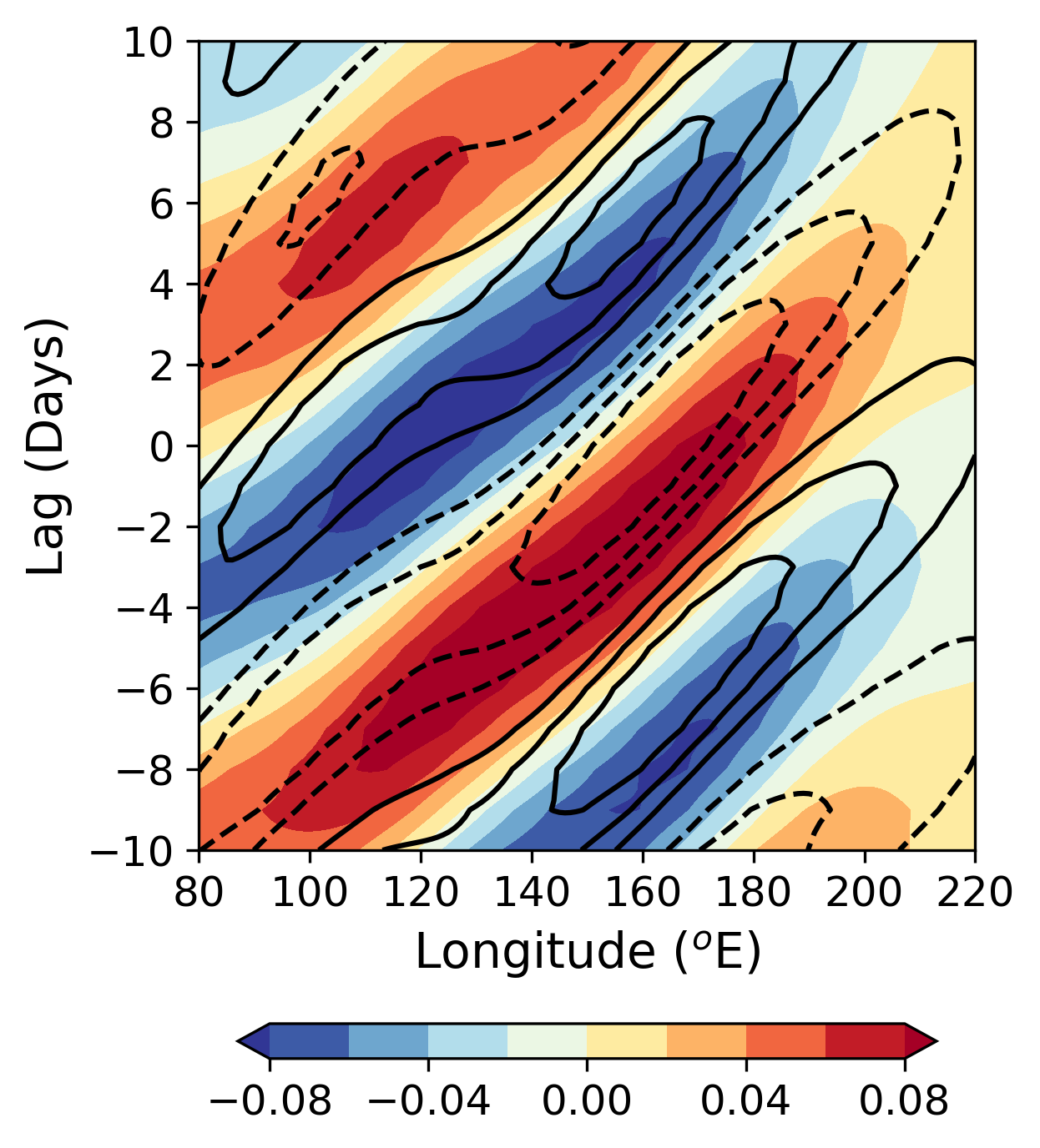}
        \caption{}
    \end{subfigure}
    \hspace{-0.5em}
     \begin{subfigure}{0.25\textwidth}
        \includegraphics[width=\linewidth]{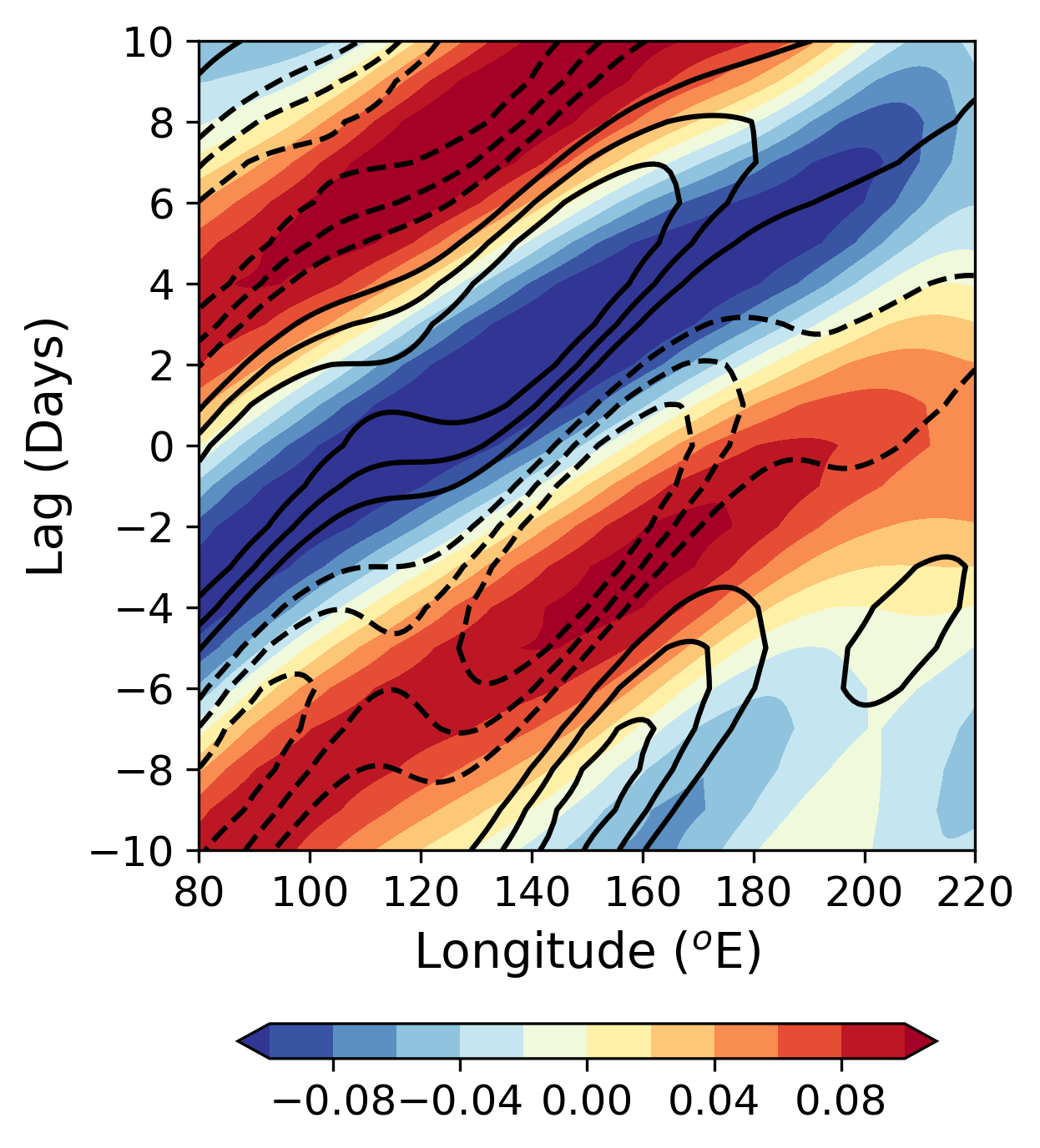}
        \caption{}
    \end{subfigure}
     \begin{subfigure}{0.25\textwidth}
        \includegraphics[width=\linewidth]{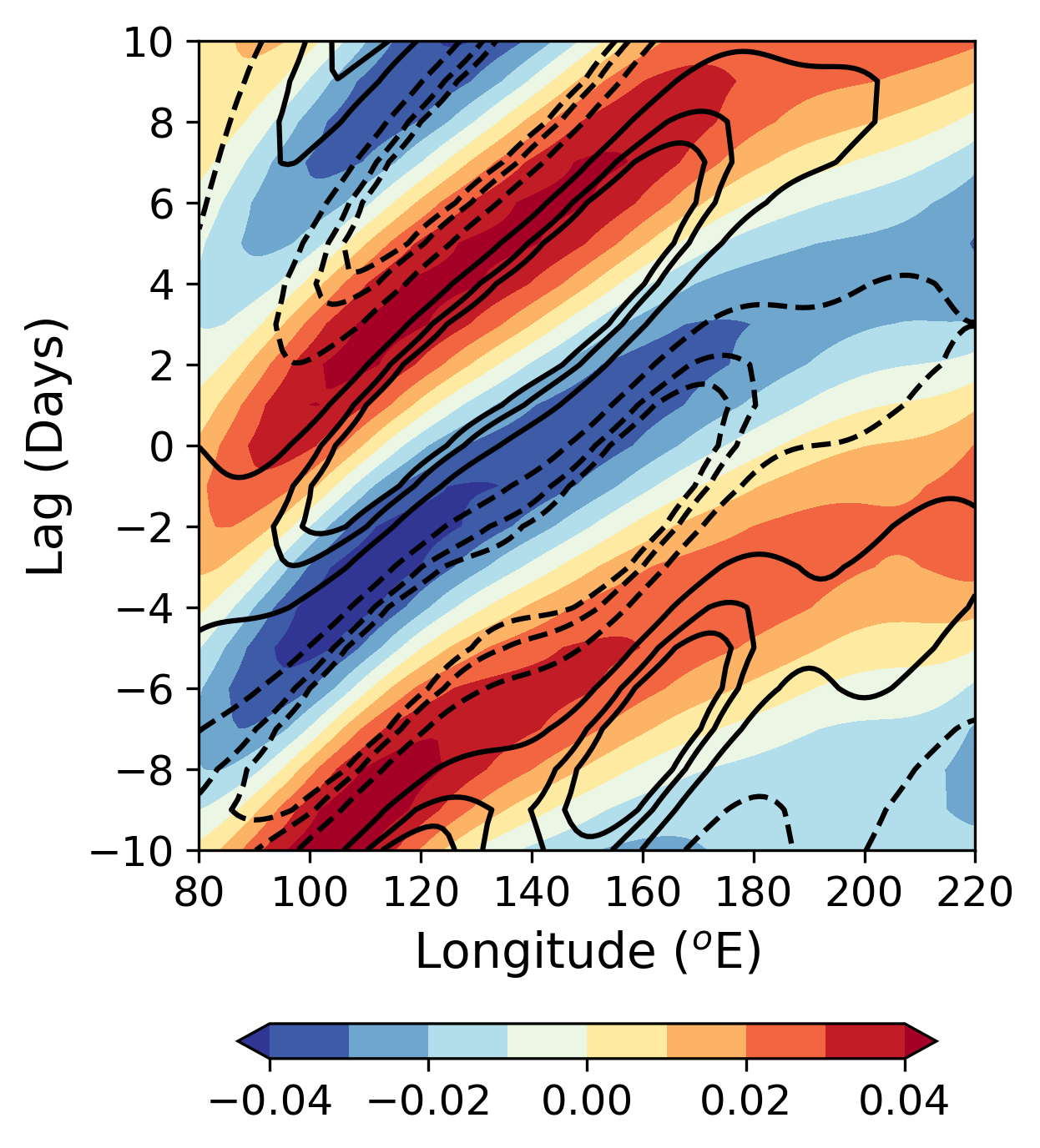}
        \caption{}
    \end{subfigure}
    \hspace{-0.5em}
    \begin{subfigure}{0.25\textwidth}
        \includegraphics[width=\linewidth]{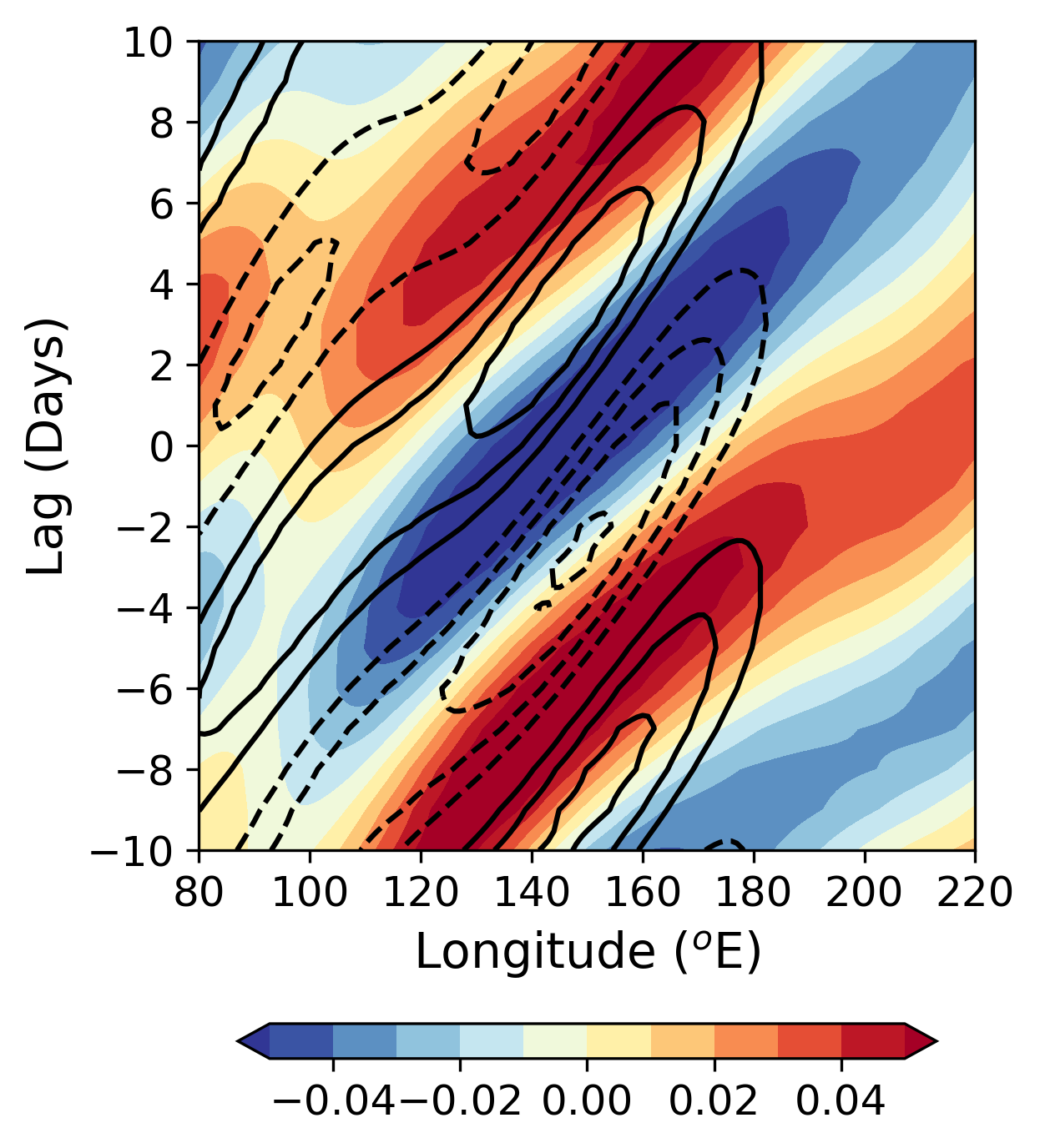}
        \caption{}
    \end{subfigure}
    \hspace{-0.5em}
    \begin{subfigure}{0.25\textwidth}
        \includegraphics[width=\linewidth]{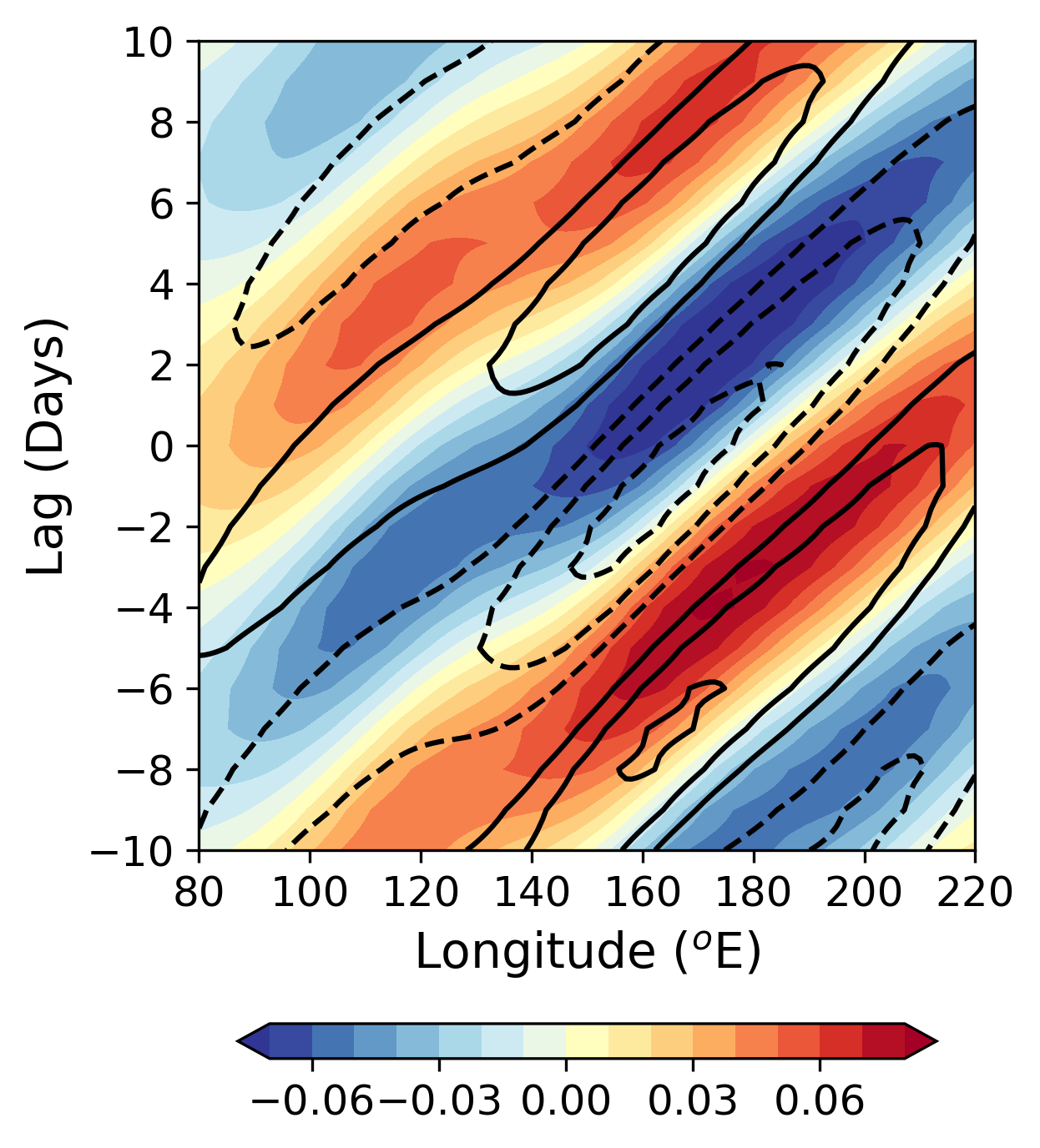}
        \caption{}
    \end{subfigure}
    \hspace{-0.5em}
     \begin{subfigure}{0.25\textwidth}
        \includegraphics[width=\linewidth]{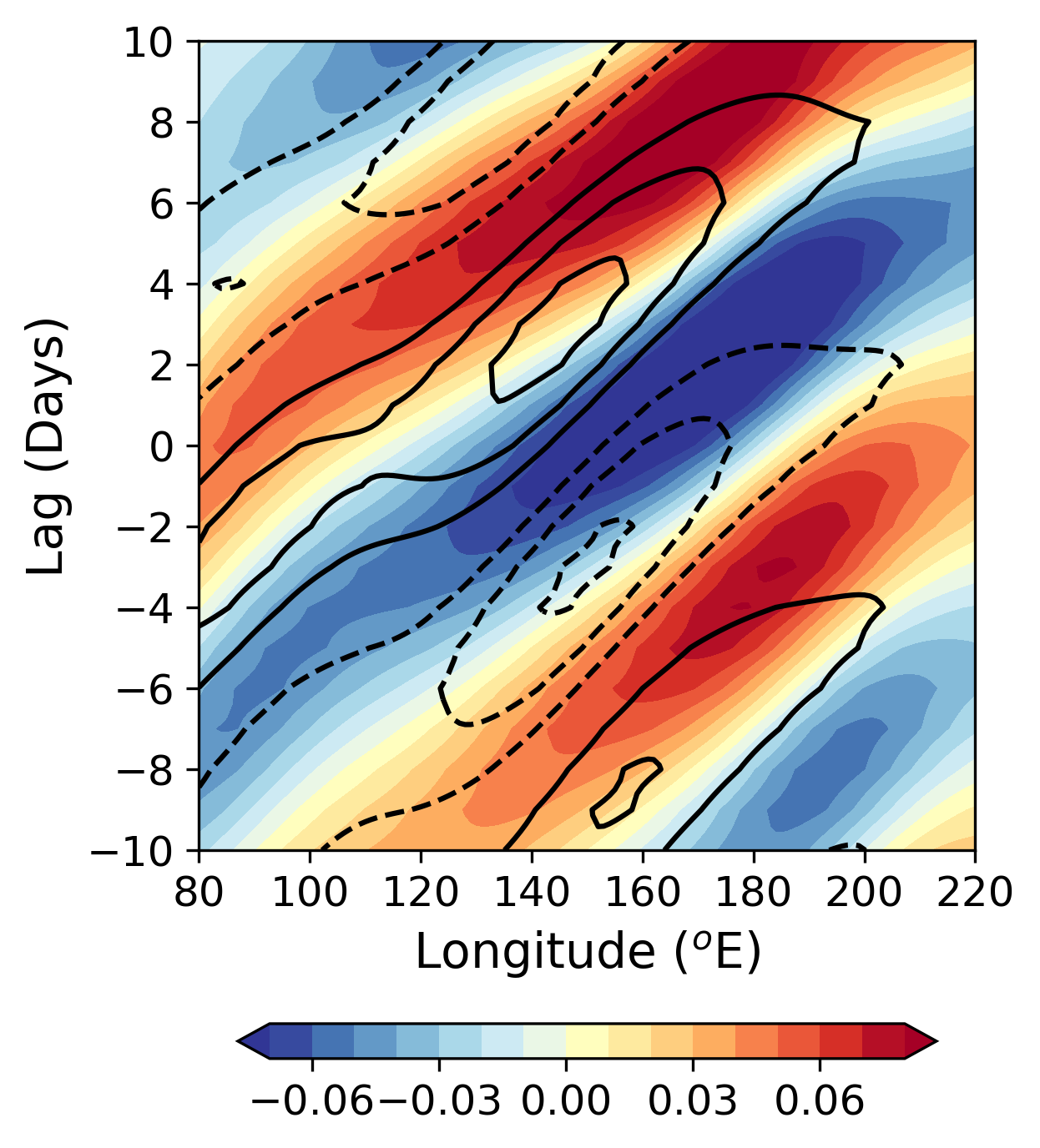}
        \caption{}
    \end{subfigure}
    \caption{Composite Hovm\"oller Diagrams for the Kelvin wave band from day $-10$ to day $+10$ at two pressure levels: 300 mb (top row; panels a–d) and 700 mb (bottom row; panels e–h). Results are shown for (a, e) PanguWeather, (b, f) GraphCast, (c, g) FourCastNet, and (d, h) Aurora. Shading indicates temperature anomalies (K), and contours represent vertical velocity anomalies (Pa/s).}
    \label{fig7}
    \end{figure}
    
\noindent To get a clear picture of the propagation of the waves that make up the composite, we construct a composite from day $-10$ to day $10$ of the identified events. Figure \ref{fig7} shows these Hovm\"oller diagrams for Kelvin wave events in the four models. In particular, we show the evolution of temperature and vertical velocity anomalies with time. From this, we estimate the phase speed of the eastward propagating wave in each model. Specifically, PanguWeather - 11.9 m/s, GraphCast - 8.6 m/s, FourCastNet - 10.5 m/s and Aurora - 10.1 m/s. While GraphCast shows a phase speed lower than typically expected values, all of them are consistent with the equivalent depths used to model these waves \citep{wk1999,straub2003,Frier}. More pertinently, the phase relation between temperature and vertical velocity anomalies in the upper and lower troposphere (in phase and downward motion leading warm anomalies, respectively) is in accord with observations \citep{straub2003} in all the four models.


\section{Rossby Wave}

\begin{figure}[H]
    \begin{subfigure}{0.5\textwidth}
        \includegraphics[width=\linewidth]{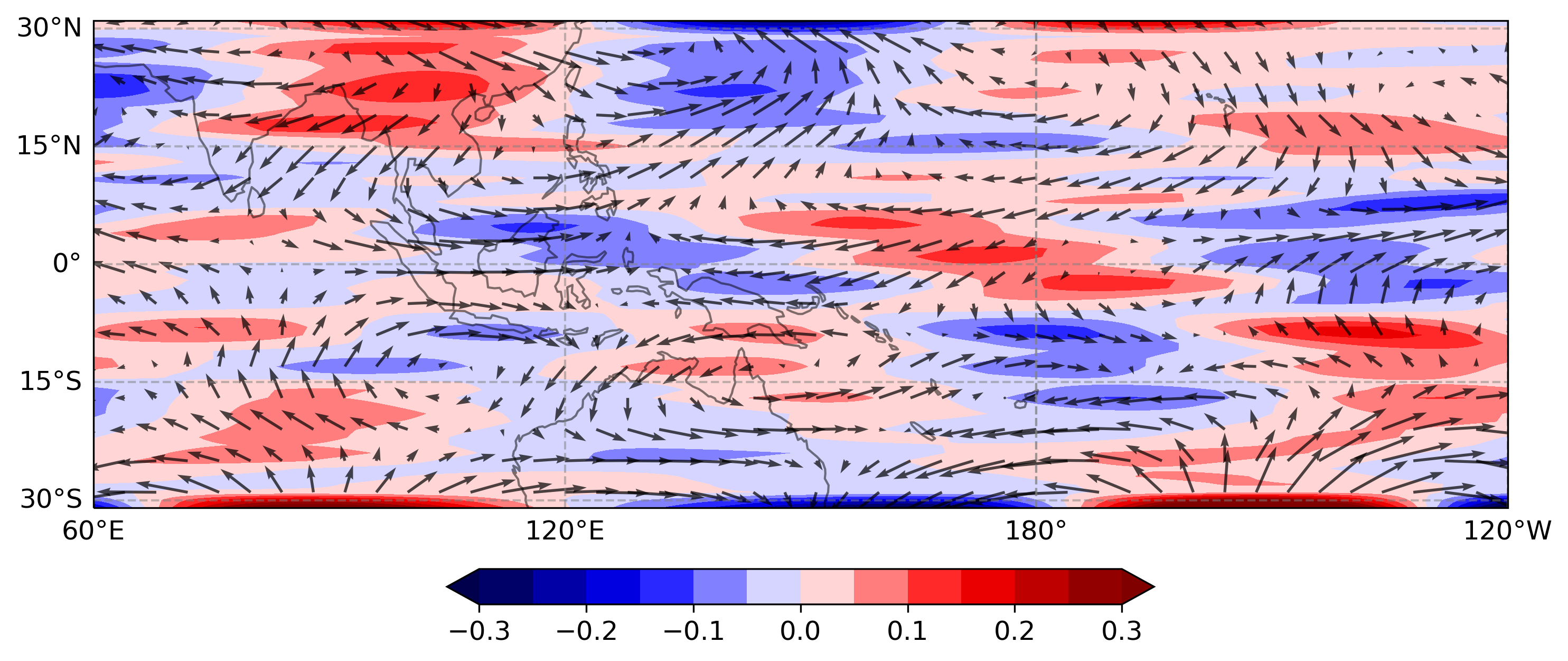}
        \caption{}
    \end{subfigure}
    \begin{subfigure}{0.5\textwidth}
        \includegraphics[width=\linewidth]{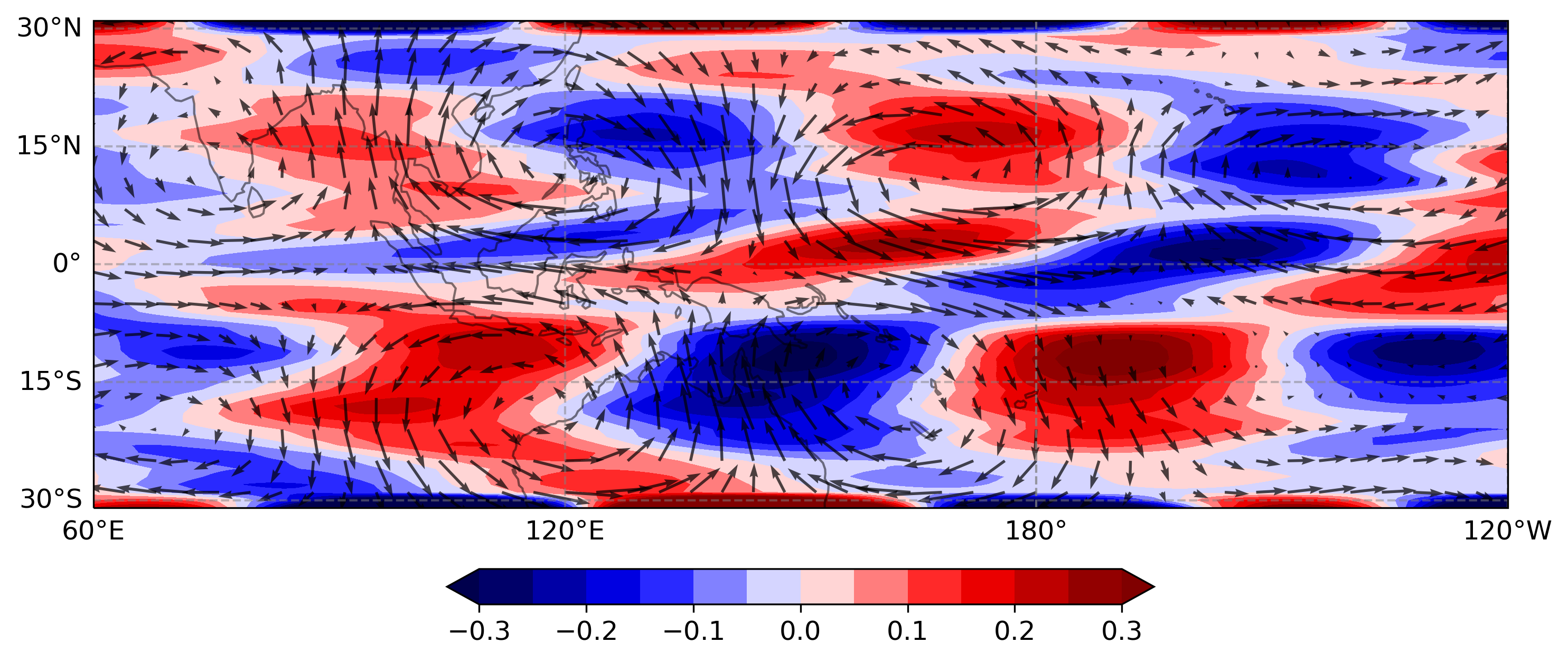}
        \caption{}
    \end{subfigure}
    \begin{subfigure}{0.49\textwidth}
        \includegraphics[width=\linewidth]{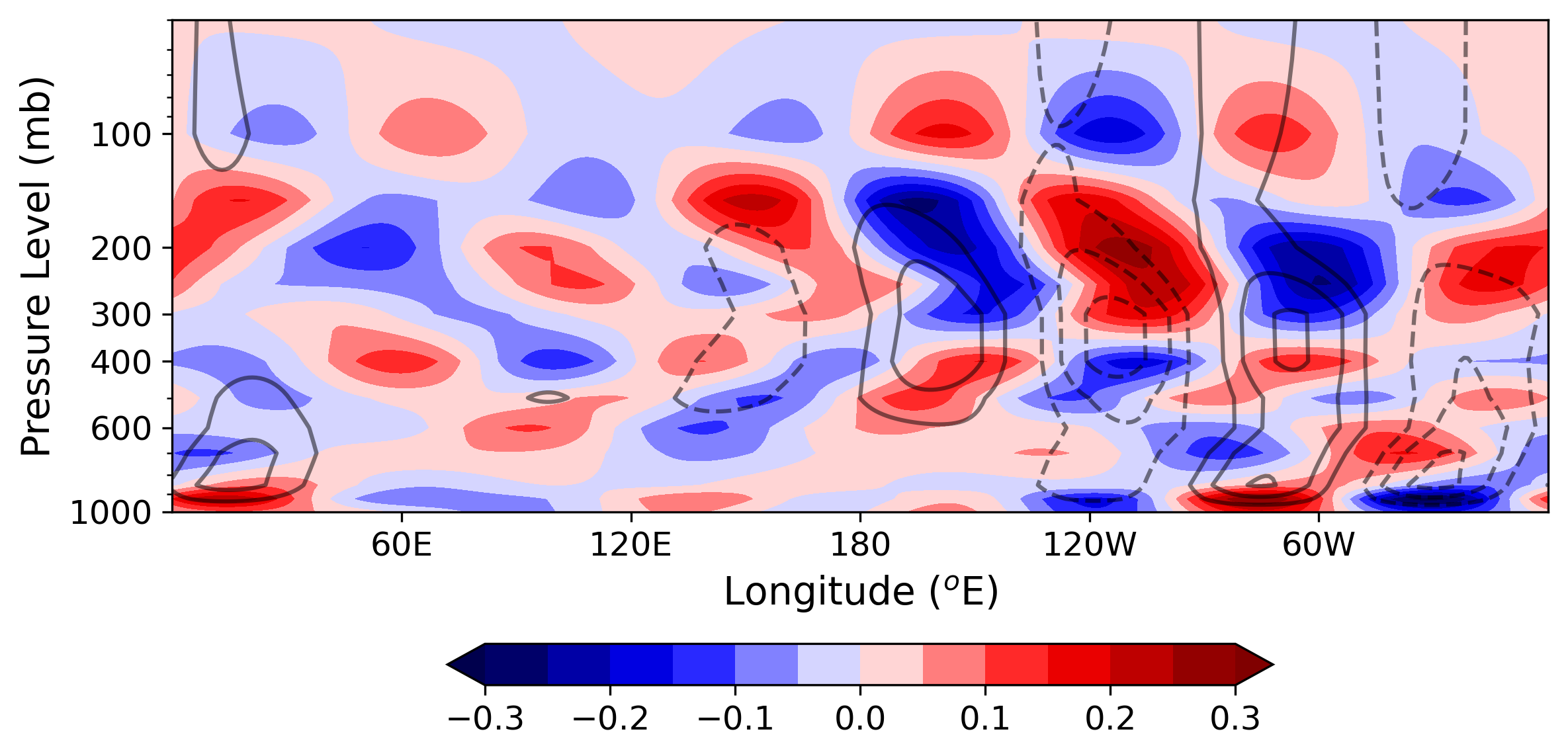}
        \caption{}
    \end{subfigure}
    \begin{subfigure}{0.49\textwidth}
        \includegraphics[width=\linewidth]{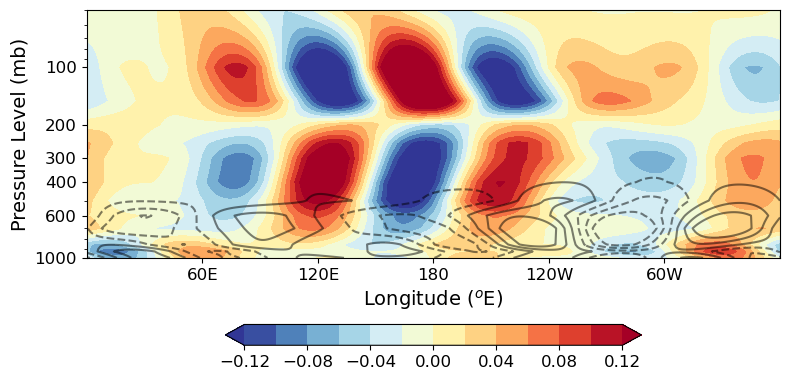}
        \caption{}
       \end{subfigure}
    \caption{Rossby Wave composite in PanguWeather. Horizontal winds (quivers) and divergence (colours, in $10^{-6}$ $s^{-1}$ ) at (a) 850 mbar, (b) 250 mbar. Vertical profiles of (c) divergence (colours, in $10^{-6}$ $s^{-1}$) with pressure velocity anomaly (contours, in Pa/s solid is positive, dashed is negative), (d) temperature anomalies (colours, in K) with specific humidity anomaly (contours, in gm/kg, solid is positive, dashed is negative).}
    \label{fig8}
\end{figure}

\begin{figure}[H]
    \begin{subfigure}{0.5\textwidth}
        \includegraphics[width=\linewidth]{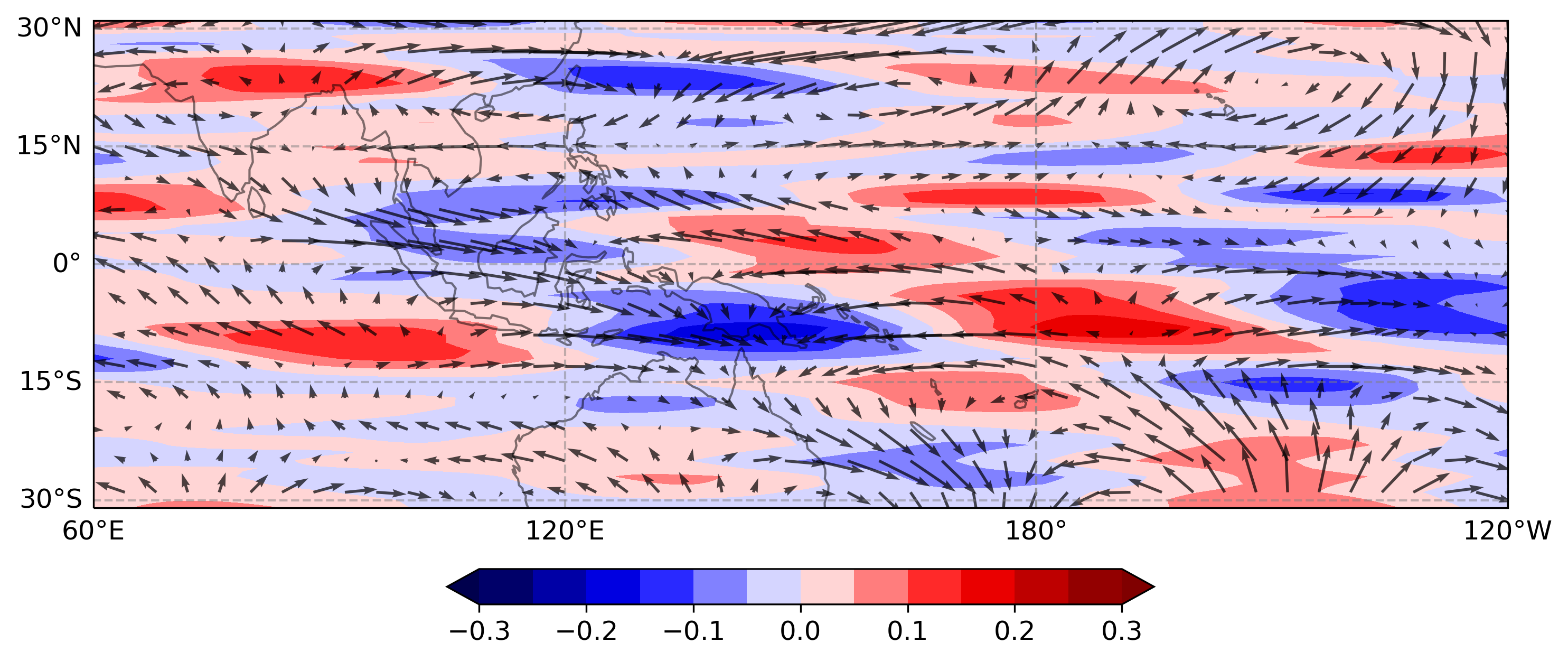}
        \caption{}
    \end{subfigure}
    \begin{subfigure}{0.5\textwidth}
        \includegraphics[width=\linewidth]{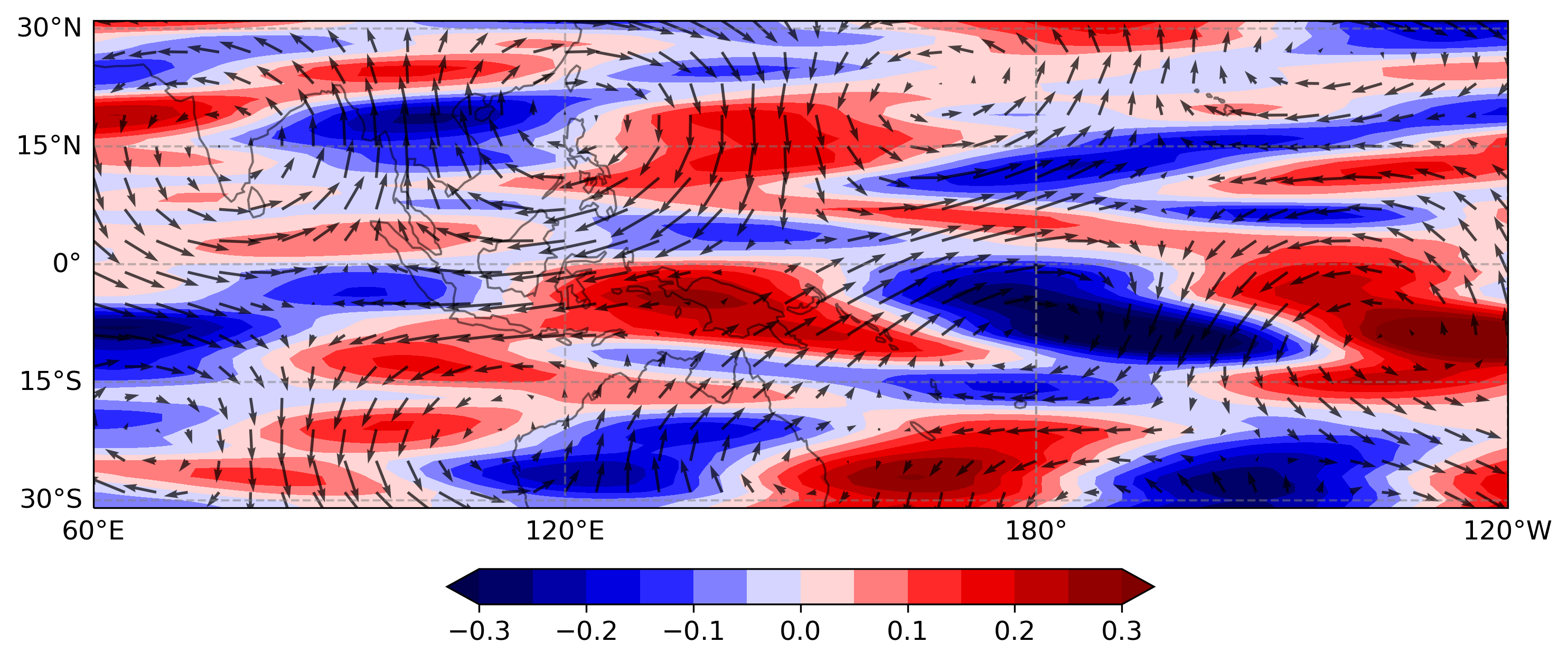}
        \caption{}
    \end{subfigure}
    \begin{subfigure}{0.49\textwidth}
        \includegraphics[width=\linewidth]{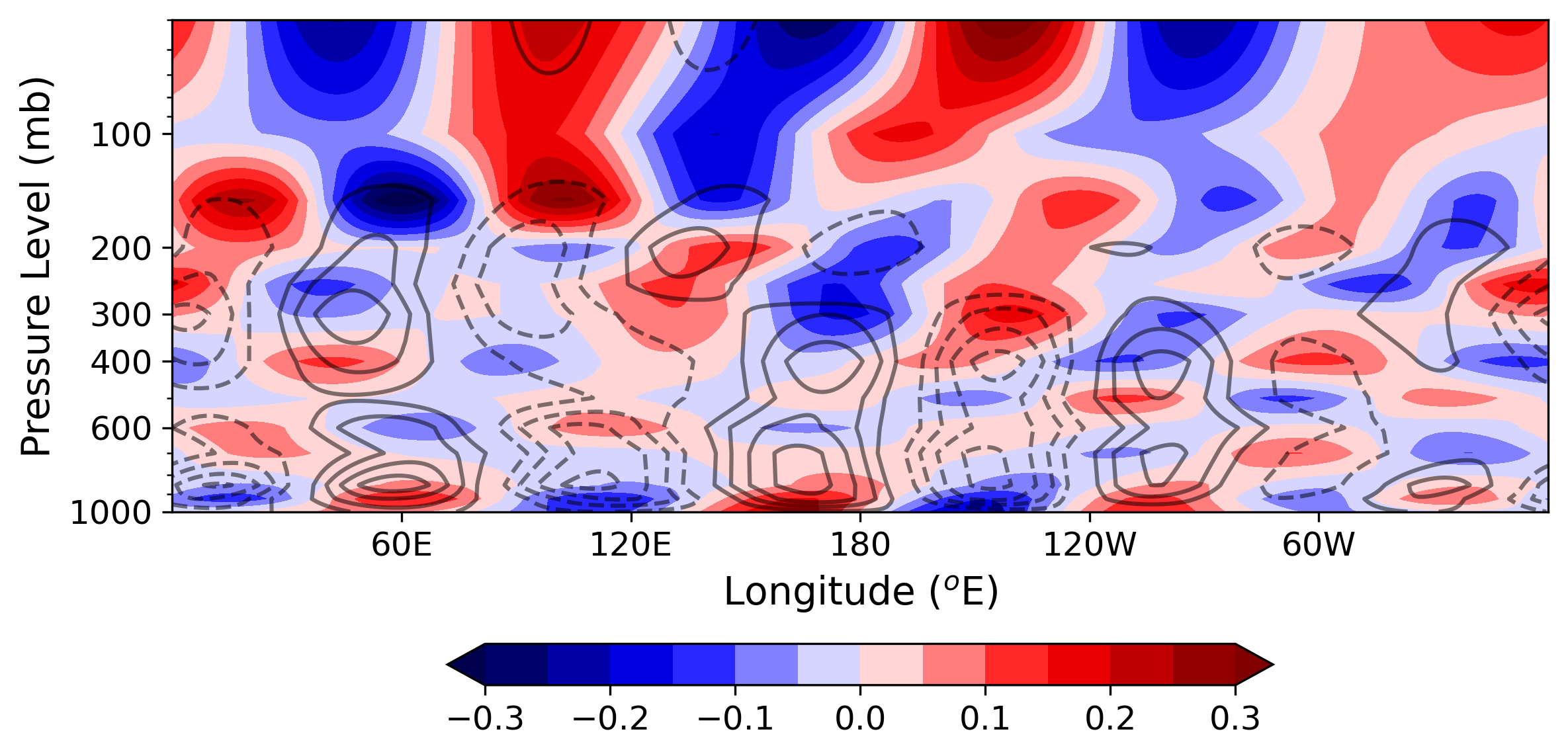}
        \caption{}
    \end{subfigure}
    \begin{subfigure}{0.49\textwidth}
        \includegraphics[width=\linewidth]{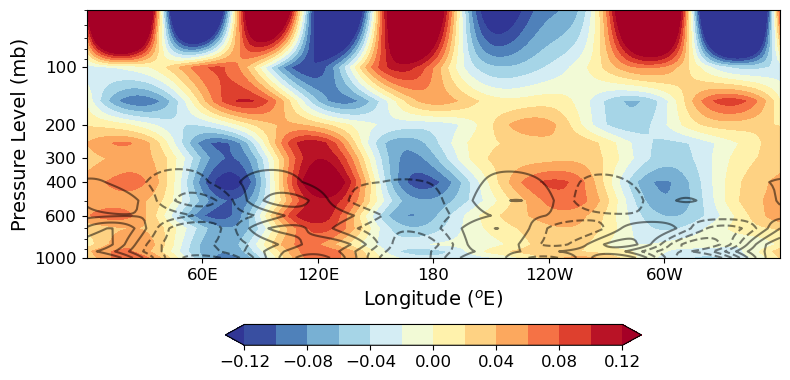}
        \caption{}
    \end{subfigure}
        \caption{Rossby Wave composite in GraphCast. Horizontal winds (quivers) and divergence (colours, in $10^{-6}$ $s^{-1}$ ) at (a) 850 mbar, (b) 250 mbar. Vertical profiles of (c) divergence (colours, in $10^{-6}$ $s^{-1}$) with pressure velocity anomaly (contours, in Pa/s solid is positive, dashed is negative), (d) temperature anomalies (colours, in K) with specific humidity anomaly (contours, in gm/kg, solid is positive, dashed is negative).}
        \label{fig9}
\end{figure}

\begin{figure}
    \begin{subfigure}{0.5\textwidth}
        \includegraphics[width=\linewidth]{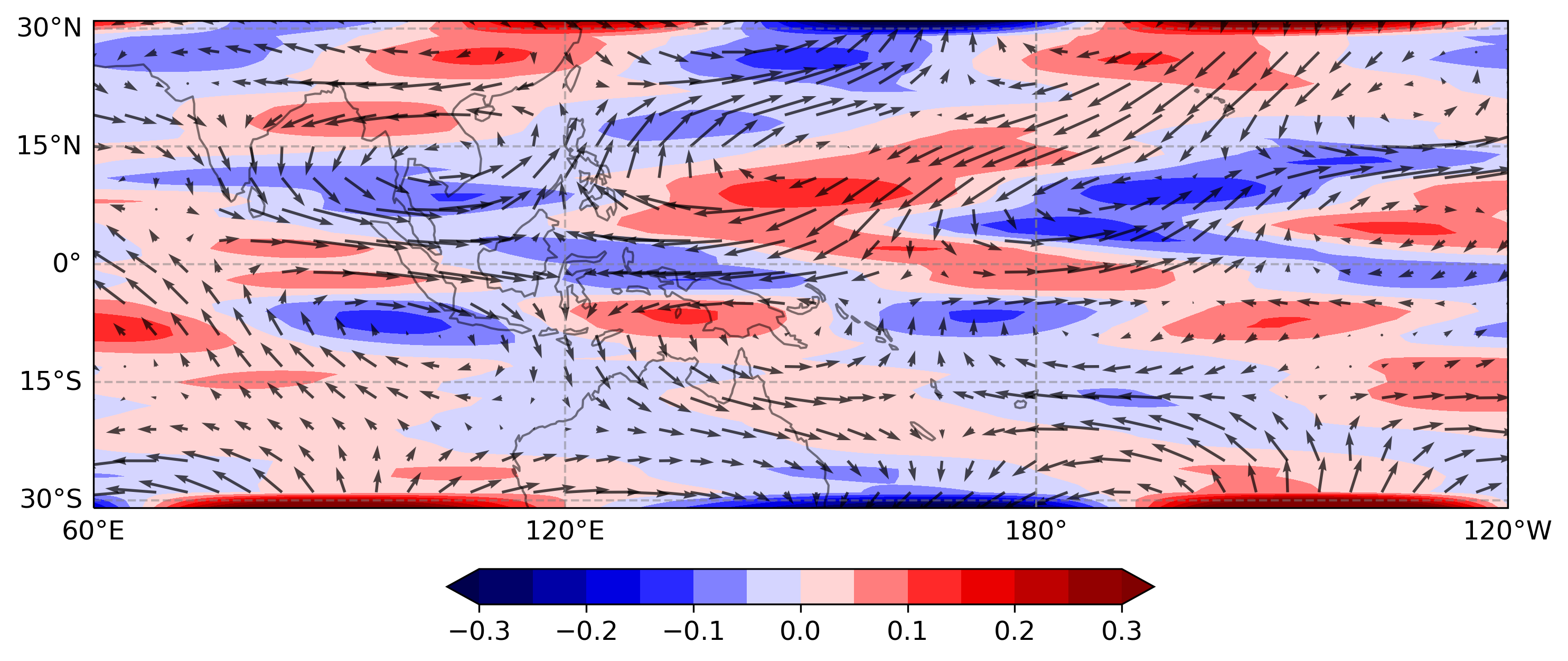}
        \caption{}
    \end{subfigure}
    \begin{subfigure}{0.5\textwidth}
        \includegraphics[width=\linewidth]{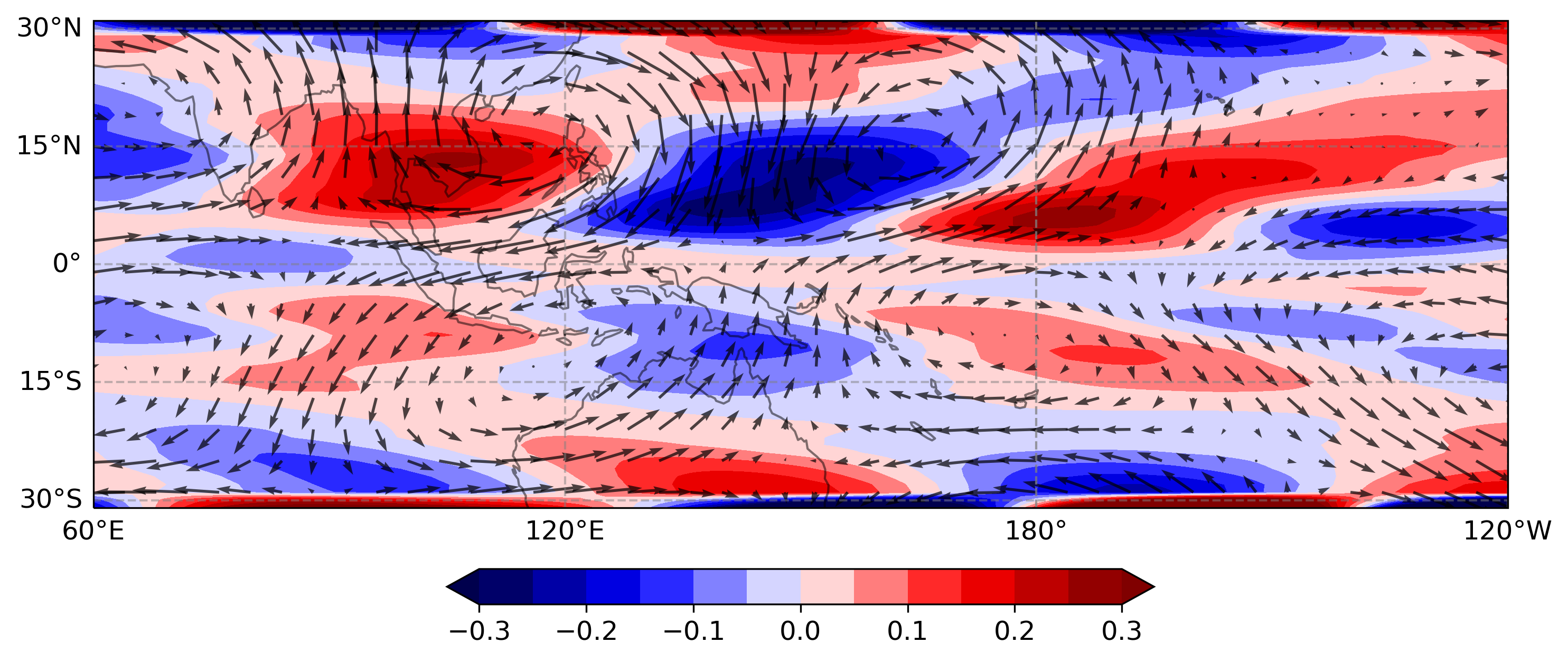}
        \caption{}
    \end{subfigure}
    \begin{subfigure}{0.49\textwidth}
        \includegraphics[width=\linewidth]{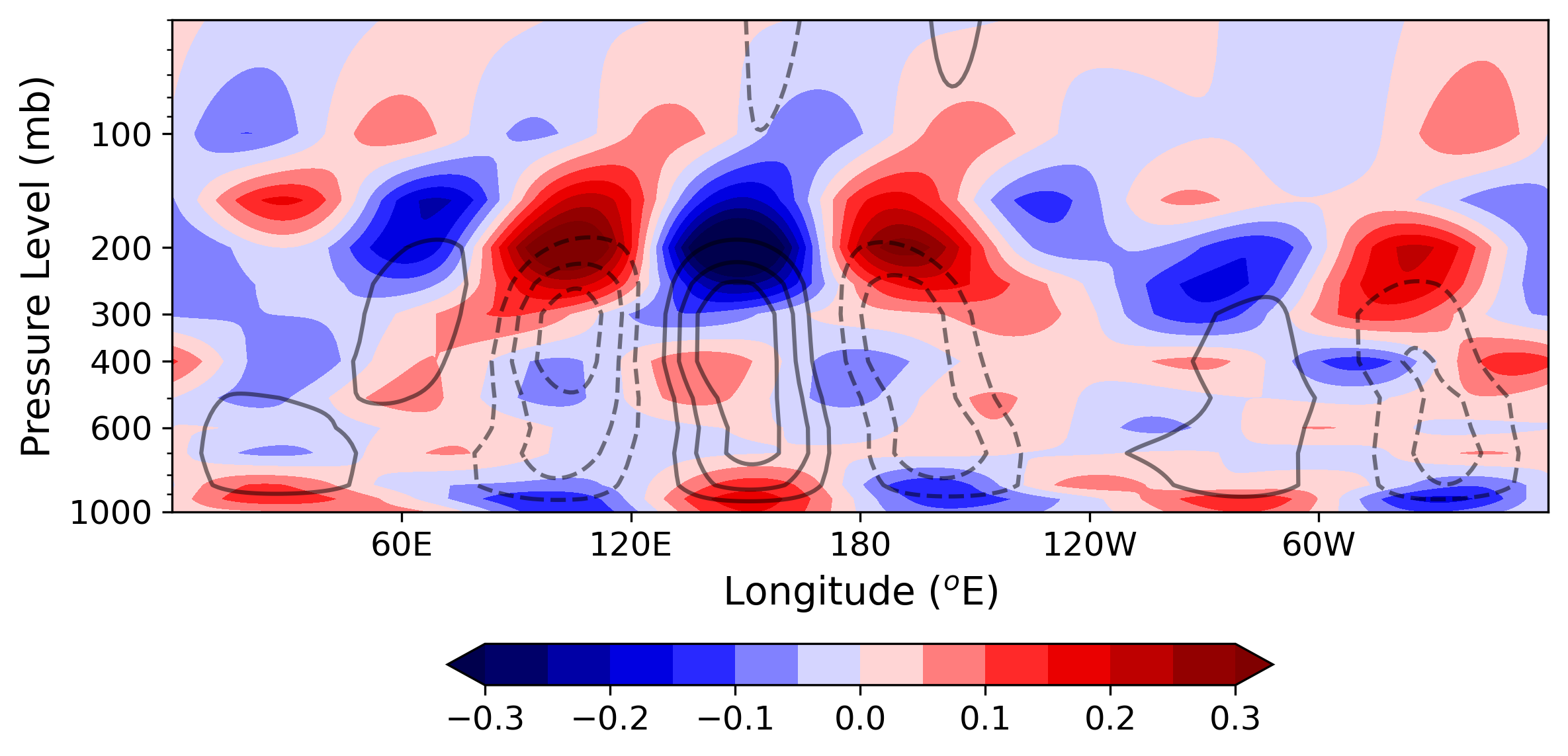}
        \caption{}
    \end{subfigure}
    \begin{subfigure}{0.49\textwidth}
        \includegraphics[width=\linewidth]{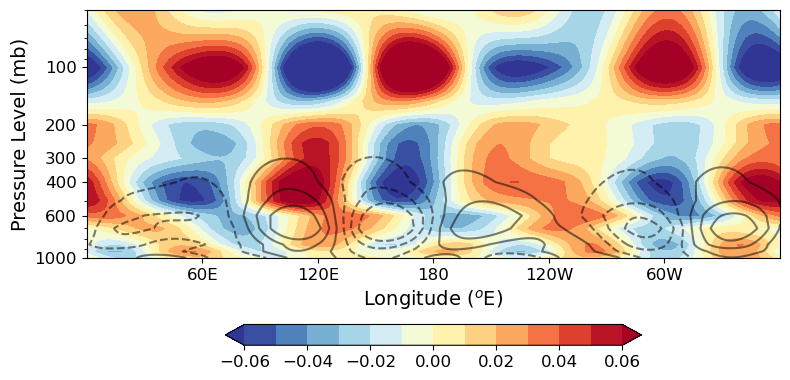}
        \caption{}
    \end{subfigure}
    \caption{Rossby Wave composite in FourCastNet. Horizontal winds (quivers) and divergence (colours, in $10^{-6}$ $s^{-1}$ ) at (a) 850 mbar, (b) 250 mbar. Vertical profiles of (c) divergence (colours, in $10^{-6}$ $s^{-1}$) with pressure velocity anomaly (contours, in Pa/s solid is positive, dashed is negative), (d) temperature anomalies (colours, in K) with specific humidity anomaly (contours, in gm/kg, solid is positive, dashed is negative).}
    \label{fig10}
\end{figure}

\begin{figure}
    \begin{subfigure}{0.5\textwidth}
        \includegraphics[width=\linewidth]{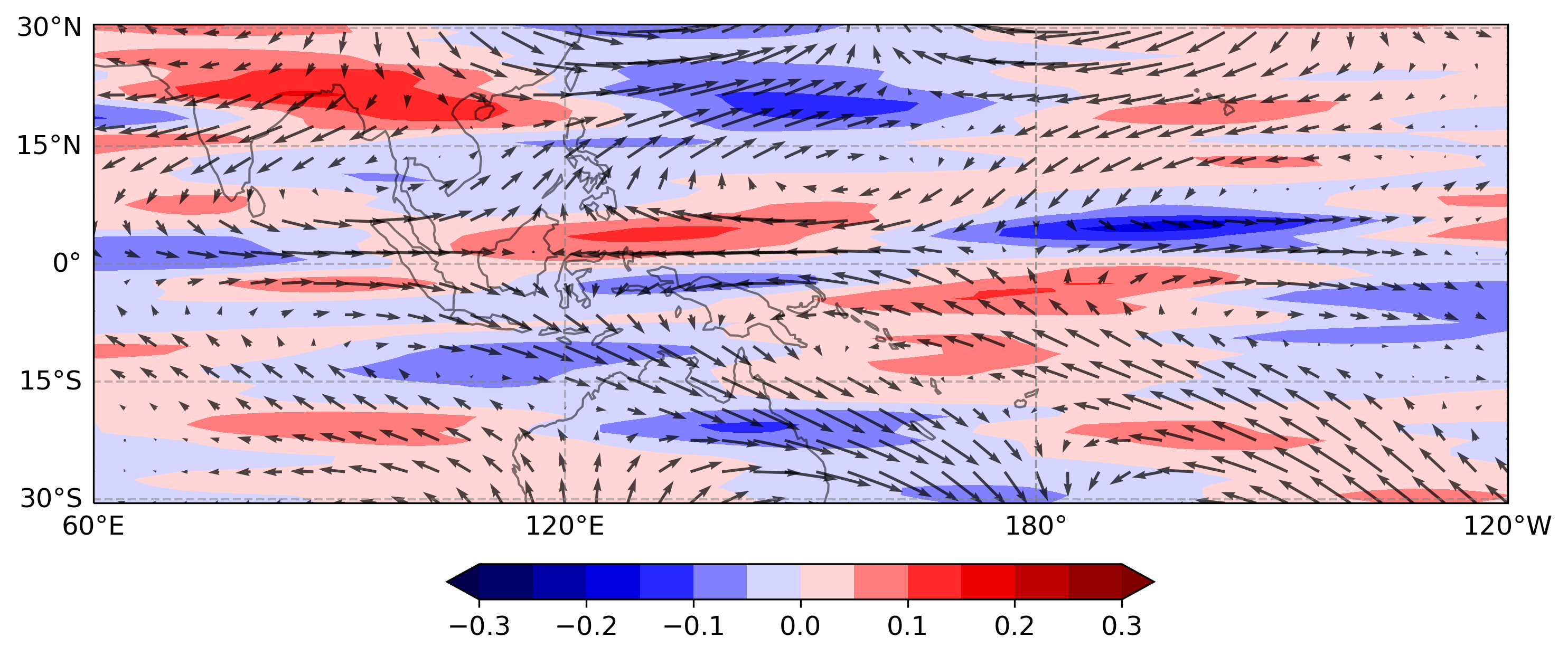}
        \caption{}
    \end{subfigure}
    \begin{subfigure}{0.5\textwidth}
        \includegraphics[width=\linewidth]{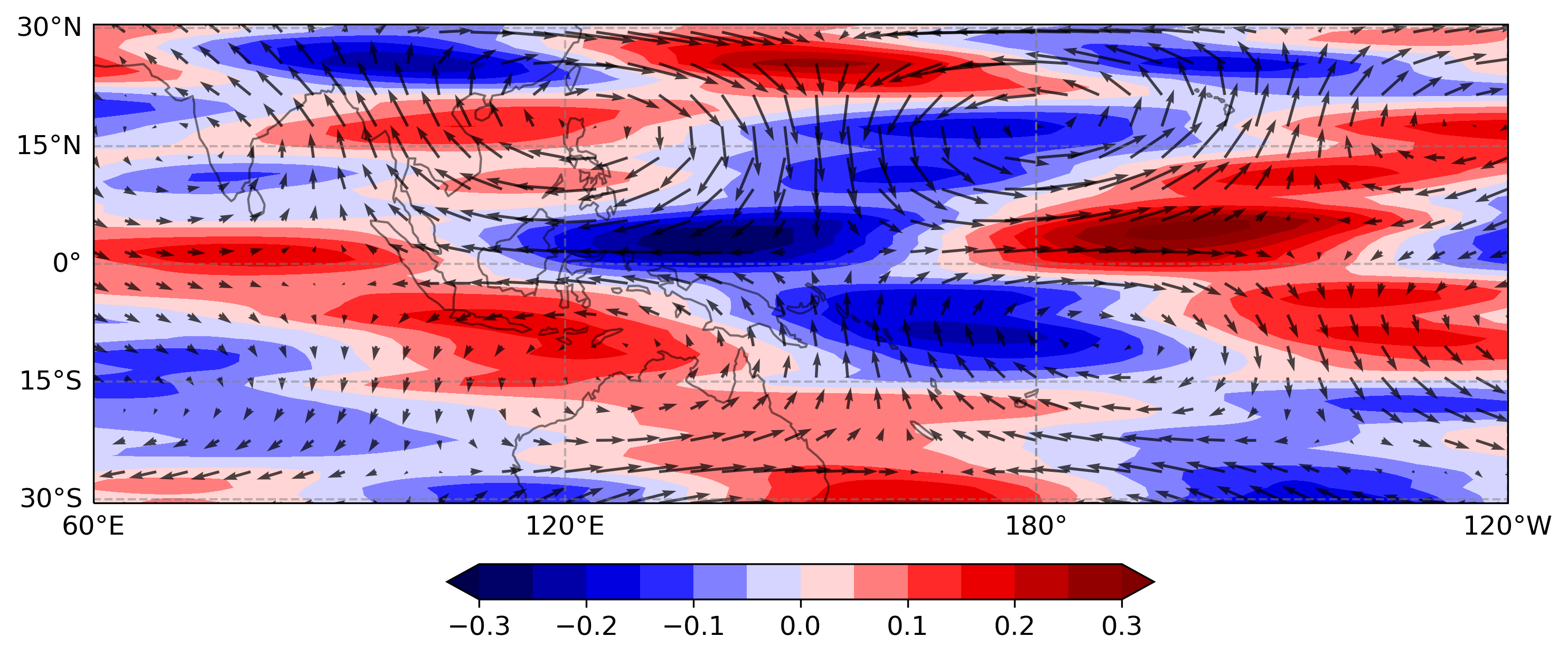}
        \caption{}
    \end{subfigure}
    \begin{subfigure}{0.49\textwidth}
        \includegraphics[width=\linewidth]{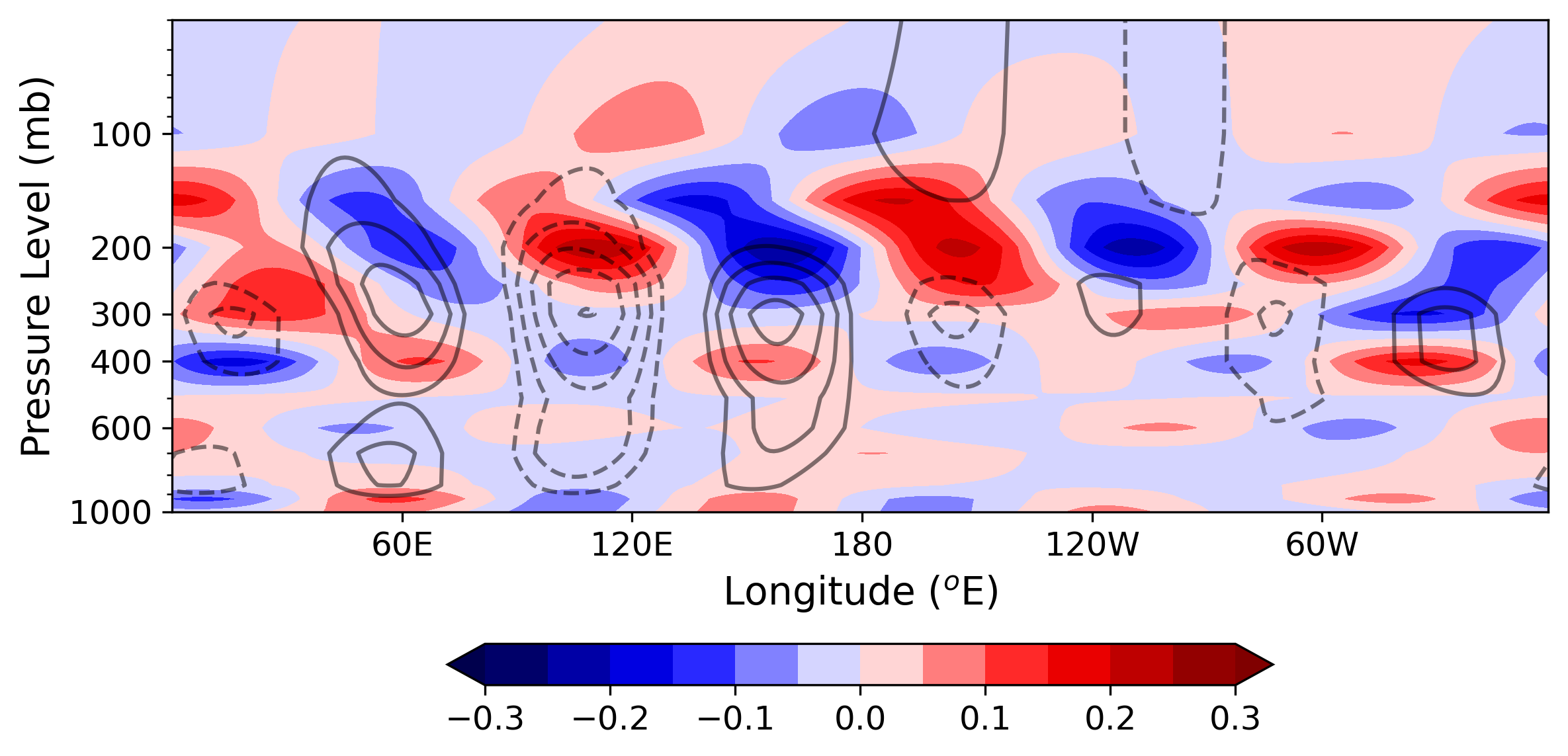}
        \caption{}
    \end{subfigure}
    \begin{subfigure}{0.49\textwidth}
        \includegraphics[width=\linewidth]{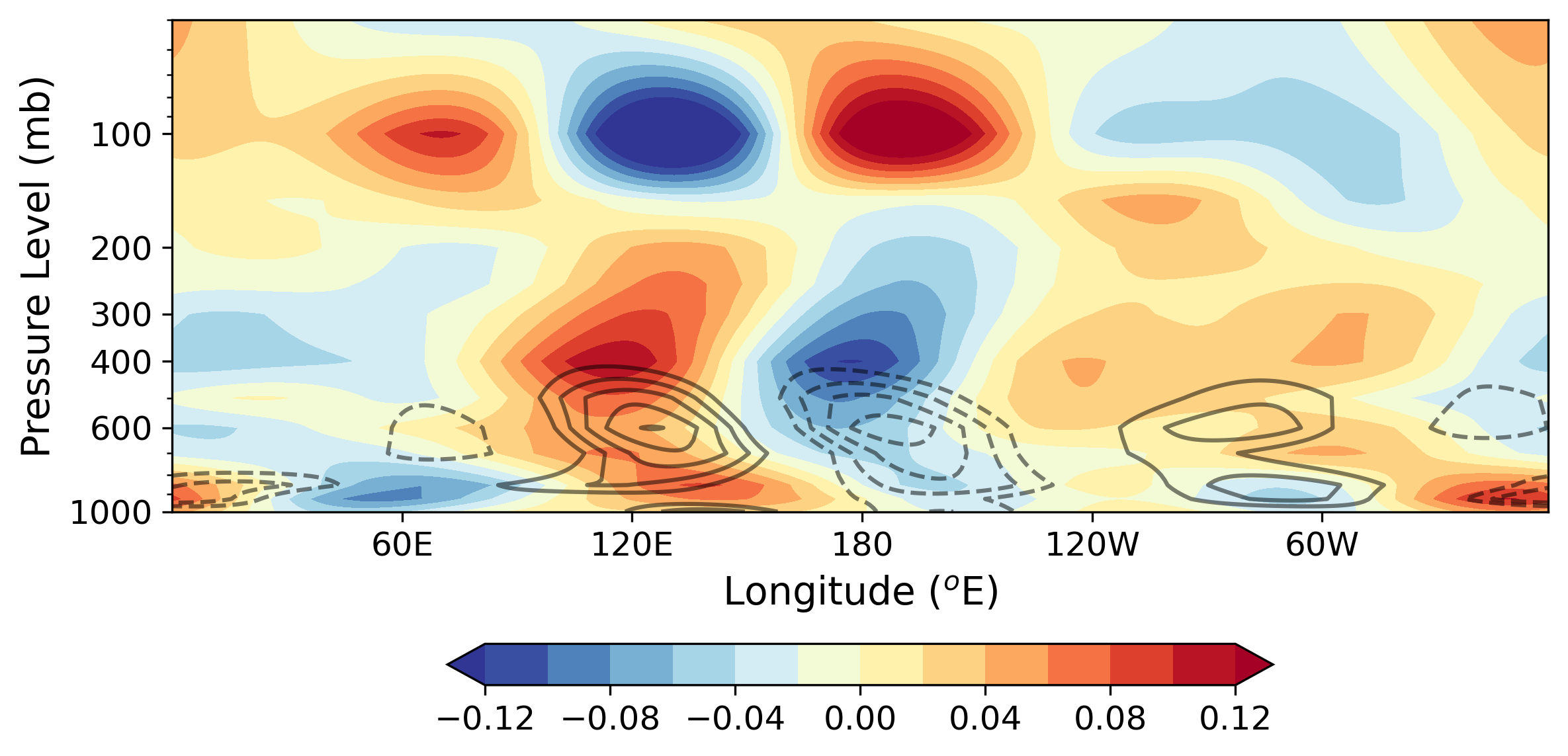}
        \caption{}
    \end{subfigure}
    \caption{Rossby Wave composite in Aurora. Horizontal winds (quivers) and divergence (colours, in $10^{-6}$ $s^{-1}$ ) at (a) 850 mbar, (b) 250 mbar. Vertical profiles of (c) divergence (colours, in $10^{-6}$ $s^{-1}$) with pressure velocity anomaly (contours, in Pa/s solid is positive, dashed is negative), (d) temperature anomalies (colours, in K) with specific humidity anomaly (contours, in gm/kg, solid is positive, dashed is negative).}
    \label{fig11}
\end{figure}

\noindent Proceeding to Rossby waves, we follow a similar strategy and analyze the structure of a composite formed by identifying events that meet the criteria mentioned in Section 2. Maps of the convergence (colors) and horizontal flow (quivers) for PanguWeather, GraphCast, FourCastNet and Aurora at 850 mbar and 250 mbar are shown in Figures \ref{fig8}a,b, \ref{fig9}a,b, \ref{fig10}a,b and \ref{fig11}a,b, respectively. Well-defined gyre pairs are evident in the 200 mbar flow in all models, particularly in the Northern Hemisphere. The upper level anticyclonic (cyclonic) gyres are associated with weak cyclonic (anticyclonic) flow at the 850 mb level. 
While the upper-level flow and gyre structures indicate the presence of Rossby waves, model performance varies considerably. PanguWeather, FourCastNet and Aurora (Figures \ref{fig8}b, \ref{fig10}b, \ref{fig11}b) correctly associate convergence (divergence) with anticyclonic (cyclonic) gyres, whereas GraphCast (Figure \ref{fig9}b) does not capture this relationship well. At the 850 mb level, FourCastNet and Aurora exhibit well defined gyres (Figures \ref{fig10}a, \ref{fig11}a), while PanguWeather and GraphCast (Figures \ref{fig8}a, \ref{fig9}a) show less organization than as noted in reanalysis \citep{kw1995,naka-taka}.

\noindent Although the models capture the correct divergence at upper levels \citep[near 250 mbar,][]{naka-taka}, they produce mid-tropospheric signals which are inconsistent with the equivalent barotropic or first baroclinic mode structures seen in observations \citep{Kiladis-rev}. This discrepancy is observed near 600 mbar in all models. Further, the lower level divergence (situated below the corresponding upper level convergence) is quite weak except in GraphCast and FourCastNet (Figures \ref{fig9}c, \ref{fig10}c). Though divergence is not particularly well represented, the vertical structure observed in the pressure velocity profiles is relatively upright, with downward (upward) motion aligning with surface divergence (convergence). This is also especially clear in GraphCast and FourCastNet (Figures \ref{fig9}c, \ref{fig10}c).

\noindent With regard to the vertical profile of divergence, the models also show a tilt in this field at some pressure levels which is not expected from reanalysis \citep{naka-taka}. PanguWeather (Figure \ref{fig8}c) shows an eastward tilt with height up to 300 mbar and a westward one above. GraphCast (Figure \ref{fig9}c) shows a large westward tilt between 60E to 180E up to 600 mbar, above which the tilt is eastward. GraphCast also shows large anomalies above 100 mbar. FourCastNet (Figure \ref{fig10}c) and Aurora (Figure \ref{fig11}c) struggle in the 400 - 600 mbar region, but represent the lower and upper troposphere reasonably well. These divergence patterns are reflected in the vertical velocity profiles. PanguWeather shows a weak and shallow structure (Figure \ref{fig8}c) , mostly confined to the upper and middle troposphere, with a deeper extension east to the region of interest. GraphCast displays two vertical motion maxima, one below 600 mbar and another above, with the lower-level structure appearing more upright (Figure \ref{fig9}c). Despite inaccuracies in the divergence fields, FourCastNet and Aurora (Figures \ref{fig10}c and \ref{fig11}c) display a coherent and vertically upright vertical velocity structure, extending smoothly from 800 mbar to 200 mbar, as is expected from reanalysis \citep{naka-taka}. 

\noindent In the vertical cross-section of temperature (Figures \ref{fig8}d, \ref{fig9}d, \ref{fig10}d, \ref{fig11}d) all of the models fail to capture the sign of the anomalies, i.e., warm (cool) anomalies are incorrectly associated with regions of upward (downward) motion. Apart from the incorrect sign, the models exhibit double maxima, first between 200 and 700 mbar, and then with a reversed-sign at and above 150 mbar. Specifically, PanguWeather shows two strong anomaly maxima, one extending from 700 mbar to 200 mbar centered around 400 mbar, and another of opposite sign near 100 mbar (Figure \ref{fig8}d). Like its divergence profile, the temperature structure exhibits a westward tilt above an eastward tilt. GraphCast also shows maxima around 300–400 mbar and opposite-signed anomaly above 200 mbar (Figure \ref{fig9}d). The profile remains relatively upright up to 200 mbar, after which it tilts westward, and similar to its divergence structure, and strong alternating anomalies appear above 100 mbar. FourCastNet shows maxima at 500 mbar and 100 mbar respectively (Figure \ref{fig10}d), with the profile exhibiting a westward tilt up to 500 mbar, transitioning to a slight eastward tilt up to 200 mbar. Aurora’s temperature structure (Figure \ref{fig11}d) resembles that of FourCastNet, with maxima at 400 mbar and 100 mbar, but with more pronounced tilts. While the tilt is not expected, the double maxima (of opposite signs) in the middle and upper troposphere have some support from reanalysis composites \citep{naka-taka} and observations \citep{Kiladis-rev}.

\noindent Interestingly, while the signs of the temperature anomalies are incorrect, all of the models capture specific humidity anomalies reasonably well. Negative (positive) humidity anomalies are collocated with regions of surface divergence (convergence), extending up to 300 mbar and peaking near the 600 mbar level which agrees with observations \citep{Kiladis-rev} and reanalysis composites \citep{naka-taka}. The specific humidity profile in PanguWeather (Figure \ref{fig8}d) appears more coherent east of the main region of interest. The anomalies show a slight eastward tilt with height and extend up to 400 mbar. Within the 60E to 180E region, while the anomaly signs are generally correct, they are not well defined. In GraphCast (Figure \ref{fig9}d), the anomalies extend from the surface up to 400 mbar with a slight westward tilt. FourCastNet (Figure \ref{fig10}d) provides the cleanest representation, with upright anomalies peaking between 600–800 mbar. In contrast, Aurora (Figure \ref{fig11}d) shows a distinct peak near 600 mbar, however it has a westward tilt in the vertical structure. Thus, while the broad specific humidity anomaly is correct, the tilts noted here are not expected from observations or reanalysis.

\vspace{0.5cm}


\begin{figure}[H]
    \begin{subfigure}{0.25\textwidth}
        \includegraphics[width=\linewidth]{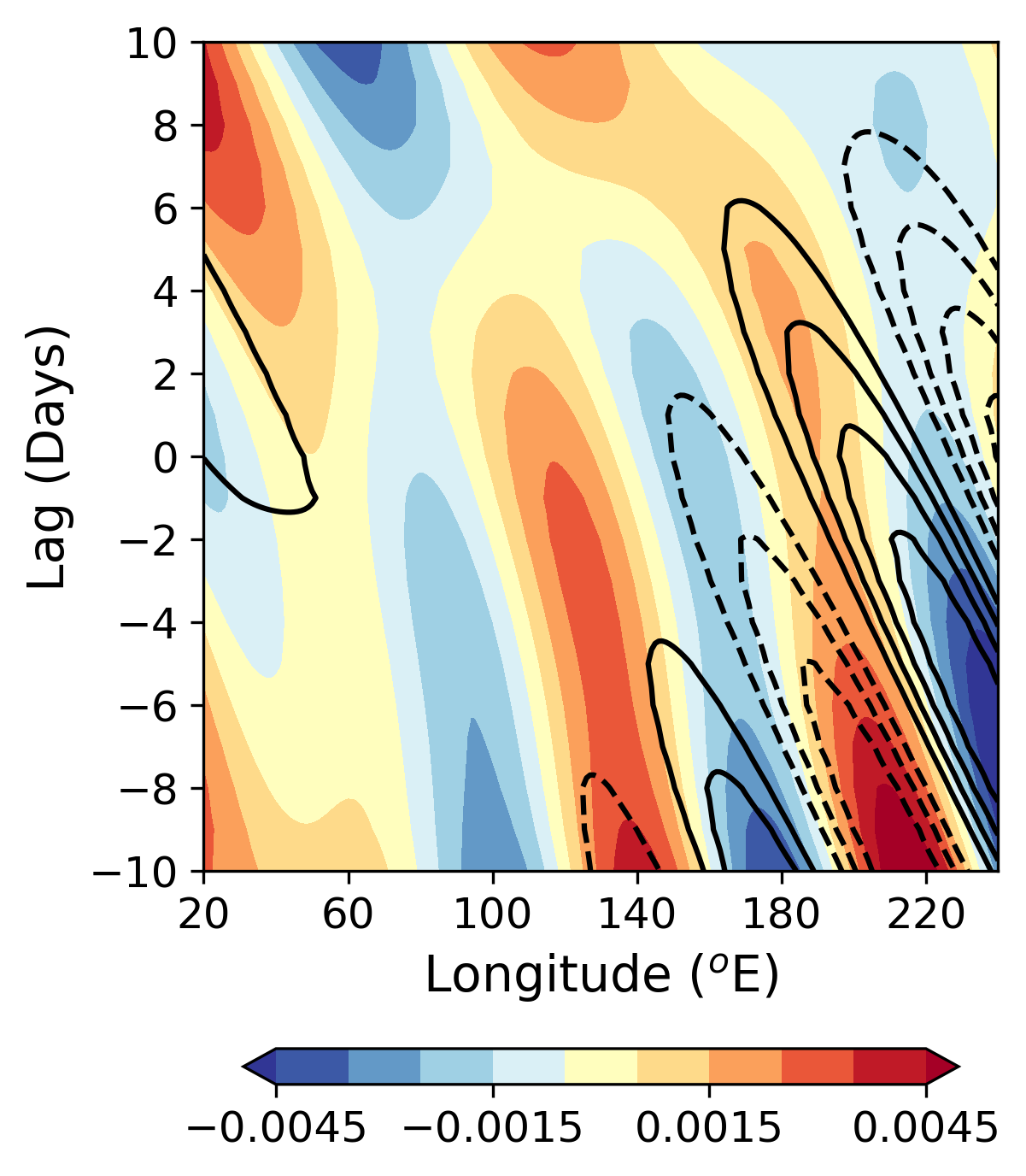}
        \caption{}
    \end{subfigure}
    \hspace{-0.5em}
    \begin{subfigure}{0.25\textwidth}
        \includegraphics[width=\linewidth]{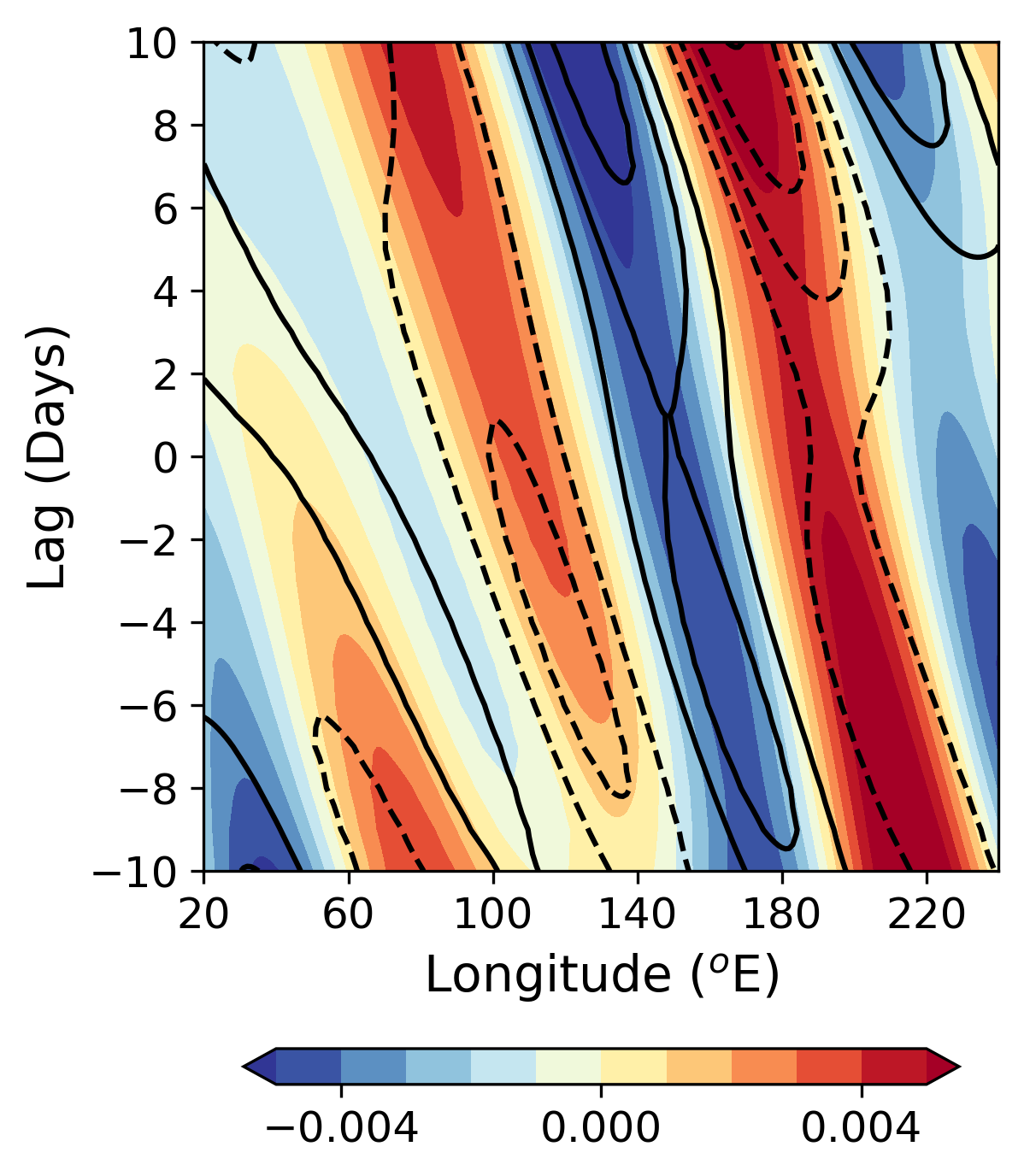}
        \caption{}
    \end{subfigure}
    \hspace{-0.5em}
    \begin{subfigure}{0.25\textwidth}
        \includegraphics[width=\linewidth]{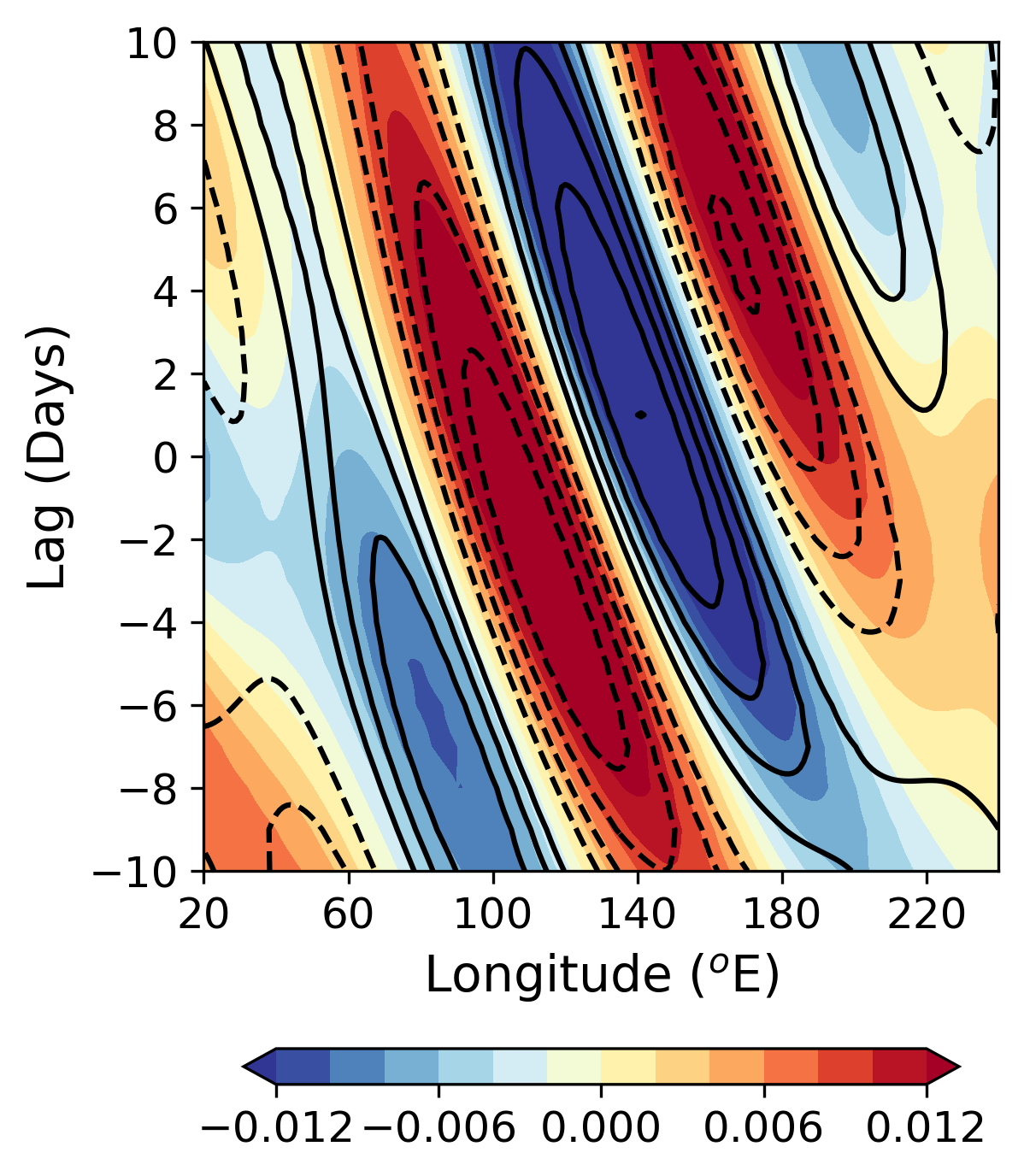}
        \caption{}
    \end{subfigure}
    \hspace{-0.5em}
     \begin{subfigure}{0.25\textwidth}
        \includegraphics[width=\linewidth]{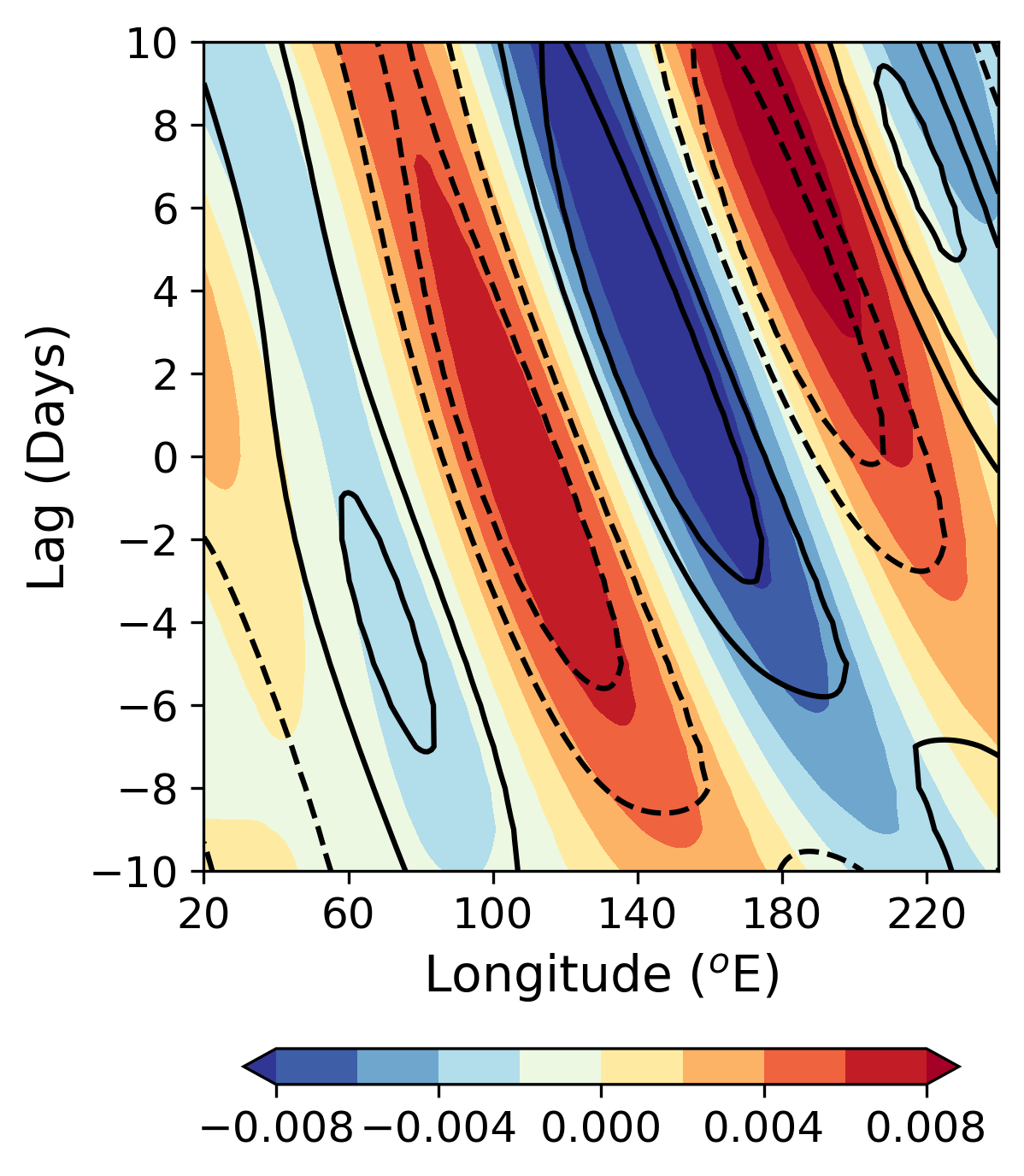}
        \caption{}
    \end{subfigure}
    \begin{subfigure}{0.25\textwidth}
        \includegraphics[width=\linewidth]{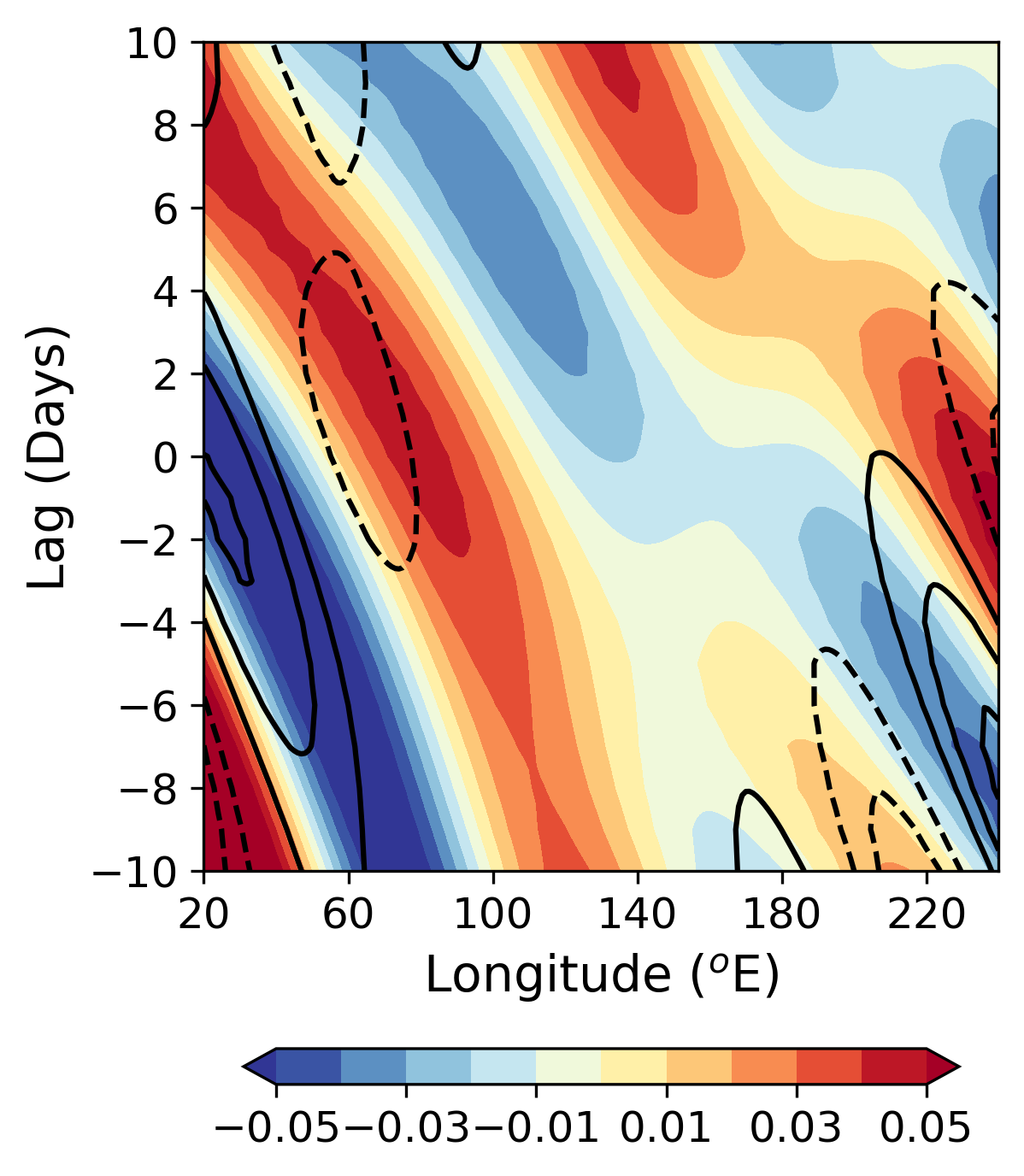}
        \caption{}
    \end{subfigure}
    \hspace{-0.5em}
    \begin{subfigure}{0.25\textwidth}
        \includegraphics[width=\linewidth]{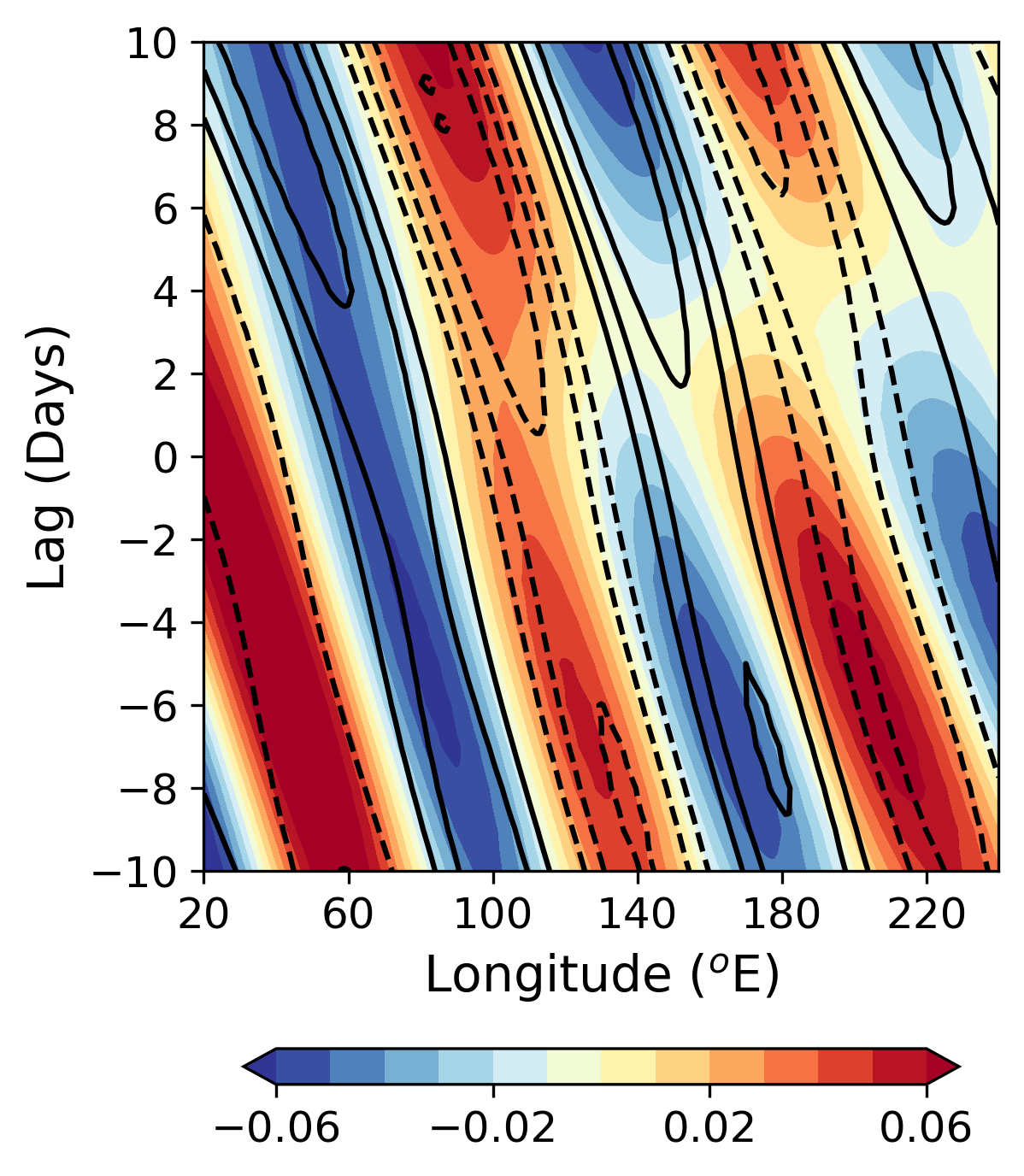}
        \caption{}
    \end{subfigure}
    \hspace{-0.5em}
    \begin{subfigure}{0.25\textwidth}
        \includegraphics[width=\linewidth]{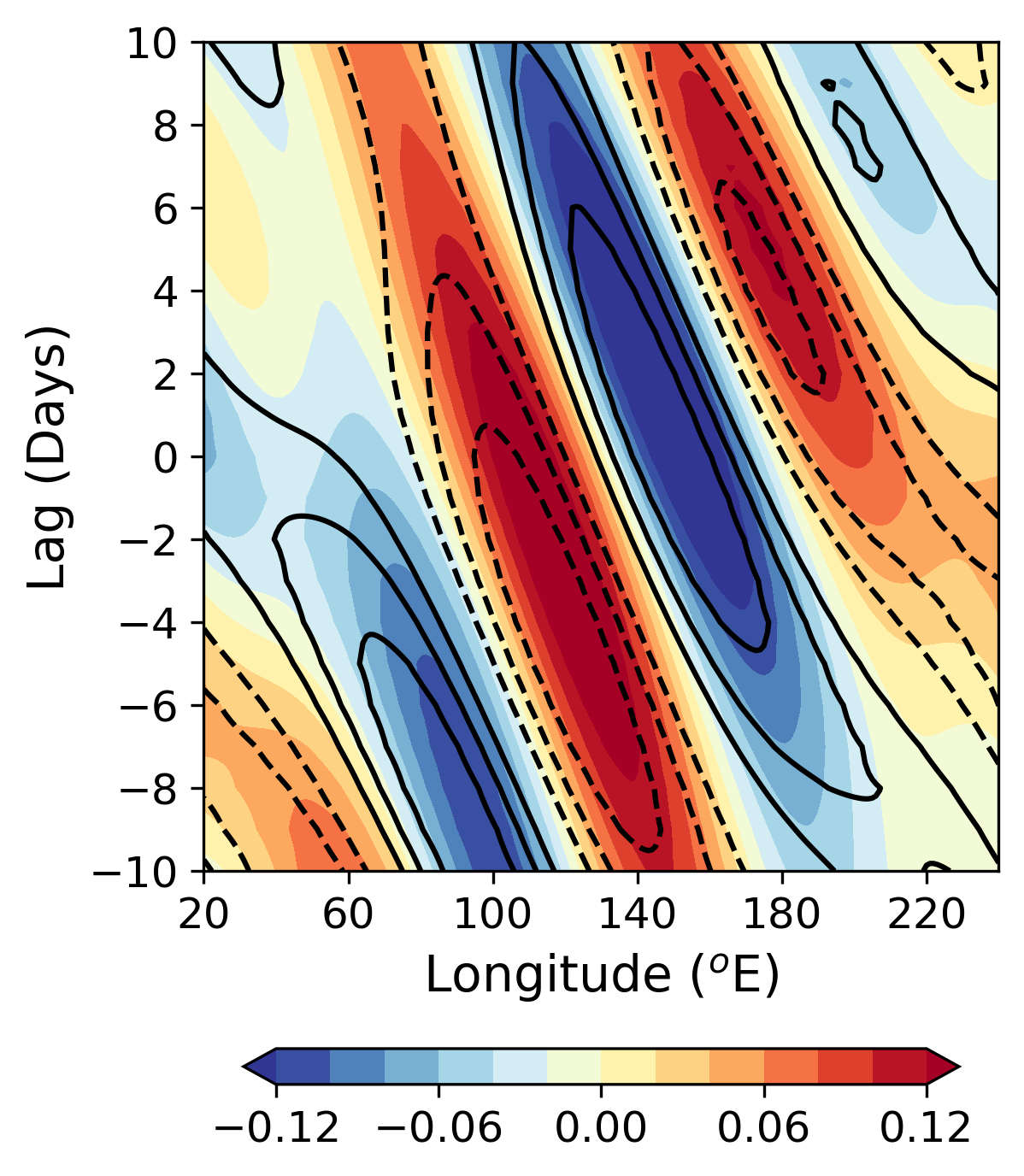}
        \caption{}
    \end{subfigure}
    \hspace{-0.5em}
     \begin{subfigure}{0.25\textwidth}
        \includegraphics[width=\linewidth]{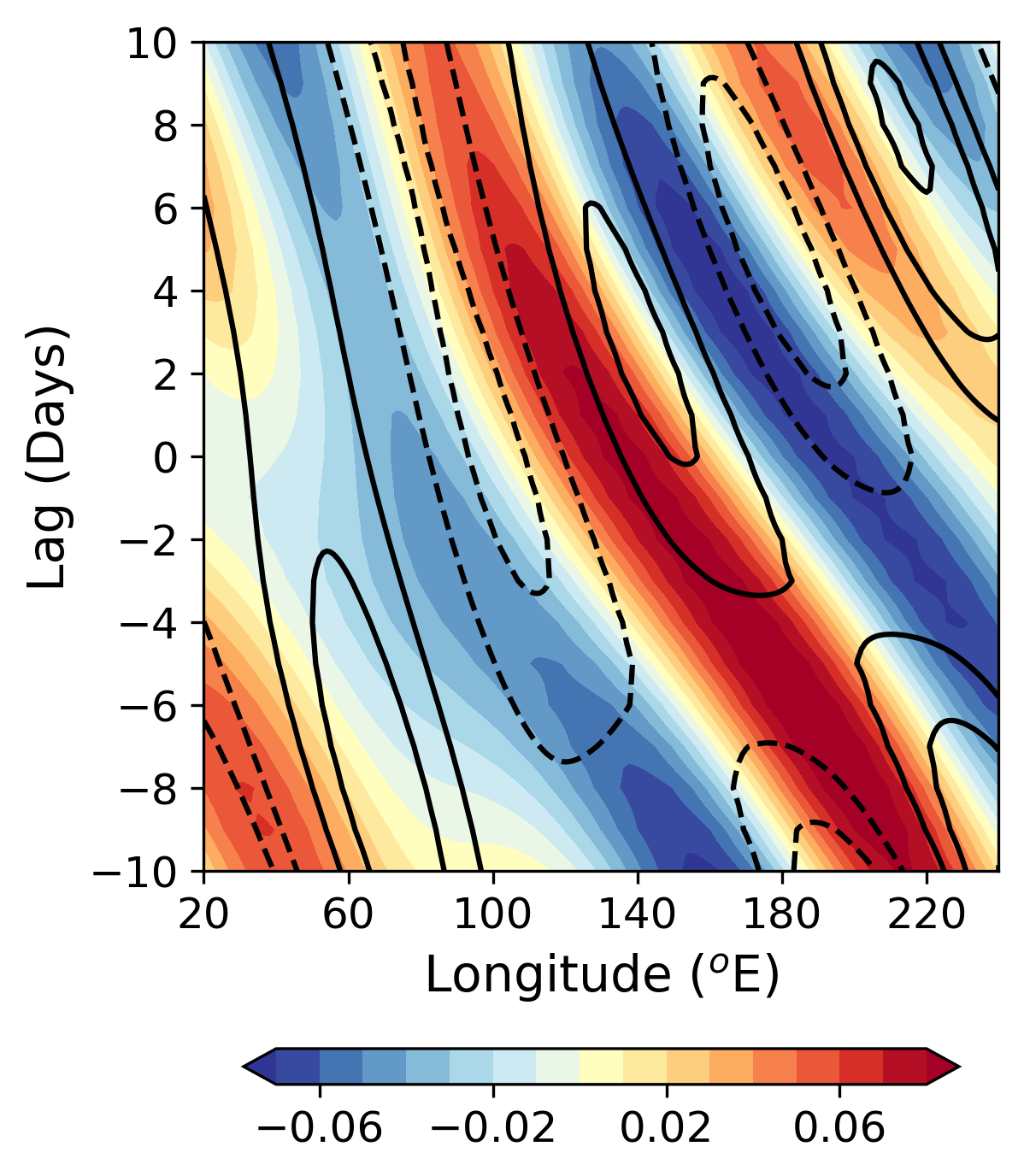}
        \caption{}
    \end{subfigure}
    \caption{Composite Hovm\"oller Diagrams for the Rossby wave band from day $-10$ to day $+10$ at two pressure levels: 300 mb (top row; panels a–d) and 700 mb (bottom row; panels e–h). Results are shown for (a, e) PanguWeather, (b, f) GraphCast, (c, g) FourCastNet, and (d, h) Aurora. Shading indicates moisture anomalies (gm/kg), and contours represent vertical velocity anomalies (Pa/s).}
    \label{fig12}
    \end{figure}

\noindent In Figure \ref{fig12}, we present a composite of the Hovm\"oller plots of moisture and vertical velocity anomalies for Rossby waves from each model. The estimated speeds are $-4.5$ m/s for PanguWeather, $-5.6$ m/s for GraphCast, $-9.6$ m/s for FourCastNet, and $-9.7$ m/s for Aurora. PanguWeather and GraphCast appear to simulate more realistic phase speeds for the equatorial Rossby waves \citep{WKW,molinari2007tropical}. While FourCastNet and Aurora better capture the overall wave structure, they are less accurate in reproducing the observed propagation speeds. Moreover, the simultaneous growth of moisture anomalies with deep vertical motion is captured by GraphCast (Figure \ref{fig12}b,f) and FourCastNet (Figure \ref{fig12}c,g), whereas PanguWeather (Figure \ref{fig12}a,c) and Aurora (Figure \ref{fig12}d,g) do not show this in phase relationship.

\section{Discussion and Conclusions}

\noindent  In this work, we have investigated the representation of tropical CCEWs in modern data-driven AI-ML models. Four models were chosen: PanguWeather, GraphCast, FourCastNet and Aurora, all of which have had great success in synoptic and medium range weather forecasts \citep{pangu2023,GraphCast,FourCastNet,aurora2025}. In particular, we study if these data-driven models have learned the basic modes of intraseasonal variability of the tropical atmosphere. Our strategy was to perform free runs of these models (outside their training period) for many months and analyze their tropical flow, temperature and humidity by performing a composite analysis of events that correspond to the various CCEWs. At the outset, a wavenumber-frequency diagram of the upper and lower level zonal winds showed heightened power in bands associated with Kelvin and Rossby waves. In fact, the equivalent depths suggested by these diagrams are also in accord with those from observations \citep{wk1999}. Other tropical waves, as well the MJO, were not clearly identifiable in these diagrams. While the higher frequency modes were not our focus, it is possible that the MJO was not adequately captured due to the length or lack of events in our simulations, or the zonal wind wasn't the most appropriate variable for this purpose. Further, the background spectrum of all models was red in character with the lower-level zonal wind decorrelation time being comparable to reanalysis, except in GraphCast where the persistence was considerable larger.

\noindent All four models performed impressively with regard to the composite structure of Kelvin waves. Not only were the lower and upper level convergence and divergence patterns captured \citep[as would be expected from the experiments using PanguWeather by][]{hakim2024}, the models showed a good agreement with observations and reanalysis based composites with regard to the vertical structure of temperature, humidity, vertical velocity and divergence anomalies. Indeed, the characteristic tilt of anomalies associated with these waves, with which many General Circulation Models have difficulty \citep{straub2010}, was well represented in these models. The distinct phase relationships between vertical velocity and temperature anomalies in the upper and lower troposphere were also captured in the models. Moreover, the phase speed of these composites was comparable to reanalysis. Of course, there were differences noted in the models. For example, the composite from PanguWeather had weak convergent and divergent flows, and was the most disorganized in terms of it's vertical profiles. GraphCast did not represent the tilt as well as FourCastNet and Aurora. Moreover, it's near surface temperature anomalies were incorrect in sign. Thus, overall, FourCastNet and Aurora appear to capture the intraseasonal Kelvin wave that is most in line with reanalysis and observations.

\noindent Proceeding to the equatorial Rossby wave, it was noted that the models had difficulty in representing it's horizontal and vertical structure. While the upper tropospheric pattern of gyres was captured well, the lower tropospheric response in most of the models was quite weak. Moreover, the vertical structure of humidity, temperature and divergence was incorrect in various ways. For example, the divergence field showed an anomalous mid-tropospheric bias and a tilt which is not in line with observations or reanalysis. Similarly, the temperature anomalies also showed an incorrect tilt with height. Specifically, PanguWeather, FourCastNet and Aurora showed a westward and eastward tilt in the lower and upper troposphere, respectively. Whereas, GraphCast had a more upright anomaly and a westward tilt above. More concerning was the fact that the temperature anomaly was of an incorrect sign and physically inconsistent with the character of the vertical velocity, i.e., negative anomalies were aligned with downward motion. Further, while the temperature anomaly was qualitatively incorrect, the moisture anomaly was by and large in accord with reanalysis and observations. Hovm\"oller diagrams showed simultaneous build-up of moisture anomalies with deep vertical velocity in both the lower and upper troposphere in the GraphCast and FourCastNet simulations. PanguWeather did not show such clear association and Aurora only followed this relationship in the lower troposphere. Finally, the phase speed of the Rossby wave was also variable among the models with PanguWeather coming closest to reanalysis based estimates. Thus, in all, while some aspects of intraseasonal large-scale Rossby wave were captured by the AI-ML models (possibly with GraphCast being most reasonable apart from the stark temperature anomaly mismatch), but overall they did not perform as well as in their representation of the Kelvin wave.

\noindent The fact that these data-driven models are able to capture characteristic features of large-scale, intraseasonal tropical Kelvin and Rossby waves is encouraging. Indeed, no specific action was taken to ensure that the models are tuned towards these modes of the atmosphere. The Kelvin wave structure in particular is very well represented in these models and longer runs with larger sample sizes would probably only improve on the composites presented here. Notably, the models have more difficulty with the Rossby wave, especially it's vertical structure. This is somewhat surprising as the rotational component has a dominant contribution to the large-scale tropical flow \citep{charney}. Given that the the large-scale equatorial Rossby wave is more rotational in character than the Kelvin wave \citep{ym2012}, one would expect it to be better captured via minimization procedures aimed at the total energy of the flow. More concerning is the physical inconsistency in some of the fields in these composites --- indeed, related issues such as deviations from geostrophic balance in forecasts have been noted in PanguWeather, FourCastNet and GraphCast at other spatio-temporal scales \citep{bonavita}. This is potentially more challenging as the data-driven models have to somehow incorporate or learn fundamental balances expected at various scales in rotating and stratified flows.

\section*{Acknowledgements}

SJ would like to thank the Prime Minister's Research Fellowship for financial support of her PhD. JS acknowledges support from the Joint Indo-Israeli DST project (DST/INT/ISR/P-40/2023). Both investigators thank the Divecha Centre for Climate Change for support. The authors also thank ECMWF and the developers of PanguWeather, GraphCast, FourCastNet and Auroa for making their models and codes freely available to the community.

\bibliographystyle{plainnat}
\bibliography{ref.bib}

\end{document}